\documentclass[%
reprint,
amsmath,amssymb,
aps,
nofootinbib]{revtex4-2}

\usepackage{graphicx}% Include figure files
\usepackage{dcolumn}% Align table columns on decimal point
\usepackage{bm}
\usepackage{bm}% bold math
\usepackage{amsfonts, cancel, bbold}
\usepackage{float}
\usepackage[caption=false]{subfig}

\usepackage{hyperref}
\usepackage{braket}
\usepackage{xcolor}
\usepackage{comment}

\hypersetup{colorlinks=true, % false: boxed links; true: colored links
allcolors=blue,        % color of links to bibliography
}

\begin{document}
\title{Opportunities for the direct manipulation of a phase-driven Andreev spin qubit}
\author{Yoan Fauvel, Julia S. Meyer and Manuel Houzet}
\affiliation{Univ. Grenoble Alpes, CEA, Grenoble INP, IRIG, Pheliqs, 38000 Grenoble, France}
\date{\today}

\begin{abstract}
In a Josephson junction, the transfer of Cooper pairs from one superconductor to the other one can be
associated with the formation of Andreev bound states. In a Josephson junction made with a semiconducting
nanowire, the spin degeneracy of these Andreev states can be broken thanks to the presence of spin-orbit coupling
and a finite phase difference between the two superconducting electrodes. The lifting of the spin degeneracy
opened the way to the realization of Andreev spin qubits that do not require the application of a large magnetic
field. So far the operation of these qubits relied on a Raman process involving two microwave tones and a
third Andreev state [M. Hays \textit{et al.}, \href{https://www.science.org/doi/abs/10.1126/science.abf0345}{Science {\bf 373}, 430 (2021)}]. Still, time-reversal preserving impurities in the
nanowire allow for spin-flip scattering processes. Here, using the formalism of scattering matrices, we show that
these processes generically couple Andreev states with opposite spins. In particular, the nonvanishing current
matrix element between them allows for the direct manipulation of phase-driven Andreev spin qubits, thereby
circumventing the use of the above-mentioned Raman process.
\end{abstract}

\maketitle

\section{\label{Intro} Introduction}
{
A Josephson junction formed via a short and narrow normal region between two superconducting leads accommodates a discrete spectrum of Andreev bound states \cite{Andreev1964,kulikMacroscopicQuantizationProximity1969a,beenakkerUniversalLimitCriticalcurrent1991d,Sauls2018}. The Kramers degeneracy of these states is lifted by the concomitance of spin-orbit coupling in the normal region and a superconducting phase bias, which breaks time-reversal symmetry \cite{chtchelkatchevAndreevQuantumDots2003b,beriSplittingAndreevLevels2008c,padurariuTheoreticalProposalSuperconducting2010b,Reynoso2012,Murani2017,parkAndreevSpinQubits2017c}. Therefore, such a Josephson junction provides a unique opportunity to realize a special kind of spin qubit, nicknamed an Andreev spin qubit \cite{chtchelkatchevAndreevQuantumDots2003b,padurariuTheoreticalProposalSuperconducting2010b,parkAndreevSpinQubits2017c}, which may not require the application of a large magnetic field to be operated, in contrast to conventional semiconductor spin qubits. Instead, the qubit operation can be performed through an ac modulation of an electrostatic gate \cite{tosiSpinOrbitSplittingAndreev2019b,metzgerCircuitQEDPhasebiasedJosephson2021c,MatuteCanadas2022,PitaVidal2023} or magnetic flux \cite{haysContinuousMonitoringTrapped2020b,haysCoherentManipulationAndreev2021}, thanks to the sensitivity of the Andreev levels to the electric potential or the phase difference, respectively. The latter option seems particularly promising. Indeed, one can anticipate a strong coupling between a Josephson junction forming part of a superconducting loop and the magnetic flux threading that loop, which is needed to set the phase difference. Experimentally so far, microwave spectroscopy allowed us to resolve the spin-splitting of Andreev levels \cite{tosiSpinOrbitSplittingAndreev2019b,haysContinuousMonitoringTrapped2020b,metzgerCircuitQEDPhasebiasedJosephson2021c,bargerbos2022spectroscopy,MatuteCanadas2022,wesdorp2022microwave}. Furthermore, the coherent manipulation of a flux-driven Andreev spin qubit was achieved thanks to a Raman process involving two microwave tones and a third Andreev level \cite{Cerrillo2021,haysCoherentManipulationAndreev2021}. These results raise the question of whether the direct manipulation of an Andreev spin qubit with a less demanding protocol, which would involve a single microwave tone, is within reach. The aim of the present work is to assess such a possibility by estimating the amplitude of the matrix element of the current operator between two states forming an Andreev spin qubit. Indeed, it is precisely this matrix element that characterizes the strength of the coupling of the qubit with an external flux drive \cite{parkAndreevSpinQubits2017c,metzgerCircuitQEDPhasebiasedJosephson2021c,Park2020,olivaresDynamicsQuasiparticleTrapping2014b}. 

To address this question, we consider the experimentally relevant situation of a Josephson junction made with a rather clean single-channel nanowire having Rashba spin-orbit coupling. When Coulomb repulsion is negligible, the ground state of the junction is even and Andreev levels are empty. However, in the odd sector a singly occupied Andreev level is long-lived \cite{Aumentado2004,zgirskiEvidenceLongLivedQuasiparticles2011a}, as a superconductor preserves parity, i.e., a second quasiparticle would be needed for the two to recombine into the even ground state. The Andreev spin qubit that we consider is formed of the two Andreev levels with the lowest energy. In the absence of spin-orbit coupling, they would form a spin-degenerate (Kramers) pair. Actually, it has been established that their spin splitting relies minimally on three ingredients: (i) an asymmetry of the velocities in opposite pseudospin bands in the nanowire (which itself necessitates a finite transverse length of the nanowire or a transverse field), (ii) a finite length of the nanowire, and (iii) a phase difference that differs from the effectively time-reversal invariant values $0$ and $\pi$. Apart from these ingredients, the amplitude of the spin-splitting is only limited by the minimum of the superconducting gap in the leads (in short junctions) and the inverse dwell time in the normal region (in long junctions) \cite{chtchelkatchevAndreevQuantumDots2003b,beriSplittingAndreevLevels2008c,haysContinuousMonitoringTrapped2020b,parkAndreevSpinQubits2017c,tosiSpinOrbitSplittingAndreev2019b}. In short, the order of magnitude of the energy spin-splitting is typically given by the same energy scale that determines the amplitude of the Josephson coupling in the considered setup. 

In the model sketched above, the matrix elements of the current operator within the Andreev spin qubit would actually vanish. Namely, the two pseudospin sectors would be completely decoupled. As a minimal model involving additional ingredients, we consider the case of a generic single scatterer located at a given position along the nanowire. We find that, for the matrix elements to take a finite value, it is necessary that the scatterer yields a finite spin-flip transmission probability with respect to the pseudospin bands in the nanowire. This is generically the case unless the scatterer possesses additional spatial (mirror) symmetries \cite{metzgerCircuitQEDPhasebiasedJosephson2021c,haysCoherentManipulationAndreev2021}. Furthermore, its location should deviate from the interfaces between the nanowire and the leads. (The matrix element also vanishes if scattering only takes place at both interfaces with the lead, but not in-between.) Our detailed study below provides the specific dependence of the energy splitting and current-operator matrix elements on the phase difference, the transmission properties of the scatterer, and its location along the nanowire. In particular, we find that the ratio between the current-operator matrix element and the energy splitting varies quadratically in the pseudo-spin band velocity asymmetry and linearly in the spin-flip transmission amplitude. The strong suppression of the current-operator matrix element does not favor the operation of the Andreev spin qubit using a flux drive if spin-orbit coupling is small. Thus, our study will contribute to identifying optimal working points, where a sufficiently strong driving may be achieved. If spin-orbit coupling is sufficiently large, the order of magnitude of the matrix element is bounded by the critical current of the junction. Thus, we do not see major challenges in operating an Andreev spin qubit in that case.

This paper is organized as follows. In Sec.~\ref{Sec::Model}, we present the model used to describe the system. In Sec.~\ref{Sec::Energy_spectrum}, we study the Andreev spectrum and determine the spin-splitting of Andreev energies. In Sec.~\ref{Curr_op}, we study the matrix elements of the current operator in the odd-parity sector. We obtain simple analytical results in the short and long junction limit, and we compare them with the numerics. Finally, we conclude in Sec.~\ref{Conclusion}.

\section{\label{Sec::Model} Model}

 In this section, we introduce the effective one-dimensional (1D) model Hamiltonian of the system. The normal part of the junction consists of a quasi-one-dimensional nanowire with Rashba spin-orbit coupling. As detailed in Refs.~\cite{Moroz1999,Governale2002,Krive2004,parkAndreevSpinQubits2017c}, the lowest subband of transverse quantization splits into two pseudospin bands with different Fermi momenta $k_{Fj}$ and different Fermi velocities $v_{j}$, where $j=1,2$, depending on the propagation direction.~\footnote{Note that it is essential to start from a higher dimensional model in order to obtain different Fermi velocities. In a strictly one-dimensional model, the Rashba spin-orbit coupling only yields a momentum shift between the two spin bands.} An example is shown in Fig.~\ref{Fig::Figure_1}. {The difference in Fermi velocities, $\delta v$, can be estimated as $\delta v \sim \alpha_R (W/L_{SO})^4(W/\lambda_F)^2$ {in the limit $W\ll L_{SO}$}, where $\alpha_R$ is the strength of the Rashba spin-orbit coupling, $L_{SO} \sim  (m\alpha_R)^{-1}$ the associated length scale, $W$ is the width of the nanowire and $\lambda_F$ is the Fermi wavelength {\cite{parkAndreevSpinQubits2017c,Martine_private}}}. In the following, we will linearize these pseudo-spin bands around the Fermi level $\mu$. The corresponding Hamiltonian $H_0$ takes the form
\begin{align}
H_0=\begin{pmatrix}H_1& 0\\0&H_2\end{pmatrix}, \quad H_j=v_j[(-1)^ji\partial_x\sigma_z-k_{Fj}],
\label{Eq::H_0}
\end{align}
in the basis $\psi=(R_1,\;L_1,\;L_2,\;R_2)^T$, where $R_j$ and $L_j$ denote right- and left-movers, respectively. Furthermore, $\sigma_z$ is a Pauli matrix in right/left space. We use units such that $\hbar=1$. Note that $R_1$ and $L_2$ ($L_1$ and $R_2$) belong to the same pseudo-spin band, see Fig.~\ref{Fig::Figure_1}. The Hamiltonian respects time-reversal symmetry (TRS), i.e.,  $\Theta H_0\Theta^{-1}=H_0$ with the time reversal operator
\begin{align}
\Theta=\begin{pmatrix} i\sigma_y\mathcal{C} & 0 \\
0 & i\sigma_y\mathcal{C} \end{pmatrix},
\label{Eq::TRS_operator}
\end{align}
where $\mathcal{C}$ denotes complex conjugation. The states $R_j$ and $L_j$ form a Kramers pair.

\begin{figure}[ht!]
\centering
\includegraphics[width=0.9\linewidth]{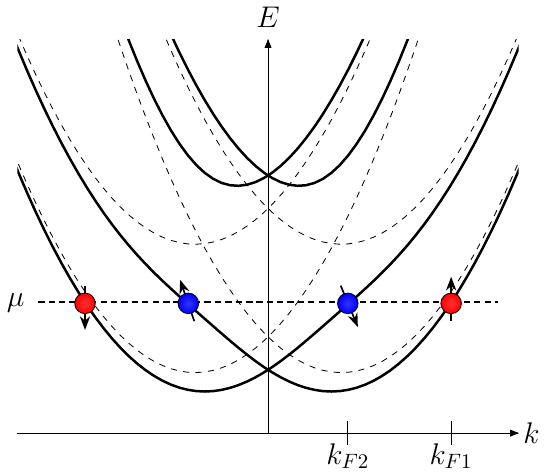}
\caption{Electron band structure of the nanowire. The dotted lines correspond to the case without coupling between the transverse subbands. The two Kramers pairs are represented by different colors. {Note that the coupling between the transverse subbands leads to a tilt between the respective spin quantization axes of the two pairs.}}
\label{Fig::Figure_1}
\end{figure}

Scattering in the normal region can be described by a Hamiltonian $H_b$ of the form
\begin{align}
H_b=\begin{pmatrix}
U_1(x) & U_3(x) \\
U_3^\dagger(x) & U_2(x)
\end{pmatrix}.
\label{Eq::H_b}
\end{align}
Assuming time-reversal invariance, { the potentials $U_{1,2}$ are constrained to be proportional to the identity, $U_{1,2}=u_{1,2}\mathbb{1}_2$ with $u_{1,2}$ real, as TRS forbids backscattering within a Kramers pair. By contrast, $U_3$ takes the form
\begin{align}
U_3&=u_0(x)\mathbb{1}_2+i\bm{u}(x)\cdot\bm{\sigma}
\end{align}
with $u_{0,x,y,z}$ real.  The diagonal terms of the block $U_3$ couple counter-propagating states within the same pseudo-spin band, whereas the off-diagonal terms of the block $U_3$} couple co-propagating states in opposite pseudo-spin bands. The latter are present, if the scattering potential possesses an asymmetry in the transverse direction \cite{metzgerCircuitQEDPhasebiasedJosephson2021c,haysCoherentManipulationAndreev2021}. In the remainder of this work, we will call these processes spin-flip scattering. As we will see, they are essential in obtaining a direct coupling between the two states of the Andreev spin qubit.

In the following, we consider a normal region of length $d$ described by the Hamiltonian $H_e=H_0+H_b$ coupled to superconductors on either side.{~\footnote{{ Here we assume that the spin-orbit coupling is unchanged under the superconductor, as also done in Ref.~\cite{parkAndreevSpinQubits2017c}. We checked that our results do not depend on this assumption. While the wavefunctions in the leads depend on the presence or absence of spin-orbit coupling, the spectrum as well as }{the matrix elements of the current operator are unchanged.}}} The superconductors induce  a pair potential with amplitude 
\begin{align}
\Delta(x) =\Delta[\theta(-x)+\theta(x-d)],
\label{Eq::def_Delta}
\end{align}
where $\theta$ is the Heaviside step function, and the phase $\phi(x)$ is equal to $-\phi/2$ and $\phi/2$ in the left and right superconductor, respectively.
Note that at this point the phase along the nanowire ($0<x<d$) is arbitrary. We will get back to the question of where the phase drop occurs at a later point.

The resulting 1D Bogoliubov-de Gennes Hamiltonian reads
\begin{align}
H_{\rm BdG}=H_e{\tau}_z+\Delta(x)[\cos\phi(x){\tau}_x-\sin\phi(x){\tau}_y],
\label{Eq::H_bdg}
\end{align}
where ${\tau}_{x,y,z}$ are Pauli matrices in particle-hole (Nambu) space, and we chose the basis $\Psi=(\psi,\Theta\psi)^T$. The particle-hole symmetry operator is given by $\mathcal{P}=-i\tau_y\Theta$ such that $\mathcal{P}H_{\rm BdG}\mathcal{P}^{-1}=-H_{\rm BdG}$. 

The Hamiltonian \eqref{Eq::H_bdg} will allow us to characterize the Andreev bound states that form in the normal region at subgap energies $|E| <\Delta$.

\section{\label{Sec::Energy_spectrum} Energy spectrum}

To derive the Andreev bound state (ABS) energy spectrum, we use the scattering formalism~\cite{beenakkerUniversalLimitCriticalcurrent1991d}. The scattering properties of the normal region are described by a scattering matrix ${\cal S}_e(E)$, { which relates incoming states $\Psi_{\text{in}}=\left(R_1^{\text{in}},\; R_2^{\text{in}},\; L_2^{\text{in}},\; L_1^{\text{in}}\right)^T$ to outgoing states $\Psi_{\text{out}}=\left(L_2^{\text{out}},\; L_1^{\text{out}},\; R_1^{\text{out}},\; R_2^{\text{out}}\right)^T$} with energy $E$ at the interfaces with the left and right superconductor, see Fig.~\ref{Fig::figure_2}. With this choice, time-reversal symmetry imposes 
\begin{align}
\mathcal{S}_e(E)=\Theta \mathcal{S}_e^\dagger (E)\Theta^{-1}
\label{Eq::restriction_TRS_scattering}
\end{align}
with the same $\Theta$ as given in Eq.~\eqref{Eq::TRS_operator}. The most general form of $\mathcal{S}_e(E)$ then reads~\cite{delplaceMagneticFieldInducedLocalization2D2012d}
\begin{align}
\!\!\mathcal{S}_e(E)=e^{i\xi(E)}
\begin{pmatrix}
r(E) & 0 & -t^*(E) & -s^*(E) \\
0 & r(E) & -s(E) & t(E) \\
t(E) & s^*(E) & r^*(E) & 0 \\
s(E) & -t^*(E) & 0 & r^*(E)
\end{pmatrix}\!.
\label{Eq::scattering_matrix}
\end{align}
Here $r(E)$ and $t(E)$ are pseudo-spin conserving reflection and transmission coefficients, while $s(E)$ describes spin-flip transmission. As pointed out before, TRS forbids spin-flip reflection. 

\begin{figure}[ht!]
\includegraphics[width=0.9\linewidth]{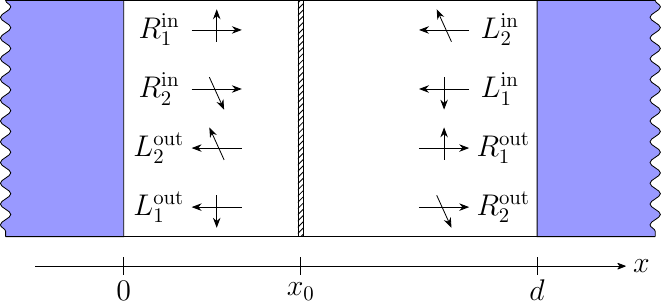}
\caption{\label{Fig::figure_2} Schematic of the scattering problem. The incoming and outgoing states are taken at both superconducting/normal interfaces. They freely propagate in the junction area of length $d$ and are scattered at the barrier at position $x_0$.}
\end{figure}

Using the particle-hole symmetry of the BdG Hamiltonian, one finds that the scattering matrix for holes is given as
\begin{align}
\mathcal{S}_h(E)=\Theta \mathcal{S}_e(-E)\Theta^{-1}={\mathcal{S}_e^\dagger(-E)}.
\label{Eq::Scattering_matrix_hole}
\end{align}
In the following, we will denote $x_e=x(E)$ and $x_h=x^*(-E)$ for $x=r,t,s$. The explicit form of the scattering amplitudes for a specific model with a $\delta$-potential will be given below.

The matrix ${\cal S}_A(E)$ describing Andreev reflection between electrons and holes at the nanowire/superconductor interfaces takes the usual form
\begin{align}
\!\!\mathcal{S}_A(E,\phi)=\alpha(E)r_A(\phi),\quad r_A(\phi)=\begin{pmatrix}
e^{i\phi/2}\sigma_0 & 0 \\
0 & e^{-i\phi/2}\sigma_0 
\end{pmatrix}
\label{Eq::matrices_andreev_reflection}
\end{align}
with $\alpha(E)=\text{exp}\left[-i\arccos(E/\Delta)\right]$. 

An ABS will form in the junction when
$\Psi_{\text{in}}={\cal S}_A(E,-\phi){\cal S}_h(E){\cal S}_A(E,\phi){\cal S}_e(E)\Psi_{\text{in}}$.
Hence, defining $M(E,\phi)=r_A(-\phi)\mathcal{S}_h(E)r_A(\phi)\mathcal{S}_e(E)$, the discrete energy spectrum of ABS is given by the roots of the secular equation~\cite{beenakkerUniversalLimitCriticalcurrent1991d},
\begin{align}
\text{Det}\left[\mathbb{1}_4-\alpha^2(E)M(E,\phi)\right]=0.
\label{Eq::Determinental_equation}
\end{align}
The solutions of \eqref{Eq::Determinental_equation} are found by diagonalizing $M(E)$, see Appendix \ref{Diag_M}, and they can be cast in the form
\begin{align}
\frac{\xi(E)-\xi(-E)}{2}+\rho\chi_\sigma(E,\phi)-\arccos\frac{E}{\Delta}-q\pi=0
\label{Eq::eq_spectre}
\end{align}
with $q\in\mathbb{Z}$ and $\sigma,\rho=\pm$. Here,
\begin{align}
&\chi_\sigma(E,\phi)=\arccos\sqrt{\frac{1+\tau\cos\left(\phi-\sigma\omega\right)+\Re\left[r_e r_h\right]}{2}}\label{Eq::chi_main_text},\\
&\omega(E) ={\text{sign}(E)}\arccos\frac{\Re\left[t_e t_h+s_e s_h\right]}{\tau},\label{Eq::omega_main_text}\\
&\tau(E)=\sqrt{(|t_e|^2+|s_e|^2)(|t_h|^2+|s_h|^2)}.
\label{Eq::val_p}
\end{align}
Note that particle-hole symmetry implies that the energy solutions of Eq.~\eqref{Eq::eq_spectre} obey $E_{q,\rho,\sigma}=-E_{-(q+1),-\rho,{-}\sigma}$. {In the following, we will thus concentrate on energies $E>0$ only. Furthermore, $E_{q,\rho,\sigma}(2\pi-\phi)=E_{q,\rho,-\sigma}(\phi)$, such that it will be sufficient to consider phases $0\leq\phi\leq\pi$. {The maximum number of ABS is set by the maximum value that $q$ can take. This value is obtained by setting $E=\Delta$ in Eq.~\eqref{Eq::eq_spectre}, leading to
\begin{align}
   q_{\rm{max}}=\left[\frac{1}{\pi}\left(\frac{\xi(\Delta)-\xi(-\Delta)}{2}+\rho\chi_\sigma(\Delta,\phi)\right)\right],
\end{align}
where $[x]$ stands for the integer part of $x$.}

A sample spectrum is shown in Fig.~\ref{Fig::Figure_3}. As can be seen, the states group into doublets labeled by an index $m\in\mathbb{N}^*$ that increases with energy. Specifically, the doublets with odd $m$ contain the states  $((m-1)/2,+,\sigma)$, whereas the doublets with even $m$ contain the states  $(m/2-1,-,\sigma)$. The energies within a doublet will then be denoted $E_{m\sigma}$.}

In the absence of backscattering, $\tau(E)=1$ and Eq.~\eqref{Eq::chi_main_text} reduces to
$$\chi_\sigma(E,\phi)=[\phi-\sigma\omega(E)]/2.$$
In this form, one can see explicitly that the spin-splitting originates from $\omega(E)$. One notices further that $\omega(E)=0$, if the scattering coefficients are energy-independent. { Note that $\sigma$ corresponds to the pseudo-spin of right-moving electrons involved in the ABS at $T=1$ and when $v_1>v_2$. In the following, we will continue to call the states $\sigma=1$ spin-up and $\sigma=-1$ spin-down even though the pseudo-spin of the ABS is not well-defined at $T\neq1$.}

For most of this paper,  we will consider a specific model with a single short-range scattering potential at a position $x_0$, i.e., in Eq.~\eqref{Eq::H_b}, we choose $U_j(x)=U_j\delta(x-x_0)$. In that case, the scattering coefficients take the following form:
\begin{align}
&\!\!\!r(E)=r\, e^{i\bar kd\tilde x_0}\!,
\enspace t(E)=t\, e^{\frac{i}{2}\delta kd}\!,\enspace s(E)=s\, e^{\frac{i}{2}\delta kd\tilde x_0}\!,\label{Eq::coef_scat_mat_s}
\end{align}
and $\xi(E)=\bar kd + \theta$. Here $\tilde x_0=2 x_0/d-1$ and $\bar k=(k_1+k_2)/2$, $\delta k=k_1-k_2$ with {$k_j=k_{Fj}+E/v_j$. Note that only the phase factors are energy-dependent. The coefficients $r$, $t$, $s$, and $\theta$ are determined by the scattering potential. Namely,
\begin{align}
&r=\frac{4 u_r^*\sqrt{v_1 v_2}}{\left\lvert\lvert u_s\rvert^2+\lvert u_r\rvert^2-\tilde{u}_1\tilde{u}_2\right\rvert},\label{r}\\
&t=-i\frac{\lvert u_s\rvert^2+\lvert u_r\rvert^2-\tilde{u}_1^*\tilde{u}_2}{\left\lvert\lvert u_s\rvert^2+\lvert u_r\rvert^2-\tilde{u}_1\tilde{u}_2\right\rvert}, \label{t}\\
&s=-i\frac{4 u_s\sqrt{v_1 v_2}}{\left\lvert\lvert u_s\rvert^2+\lvert u_r\rvert^2-\tilde{u}_1\tilde{u}_2\right\rvert},\label{s}
\end{align}
and $\theta=-{\pi}/{2}-\arg\left[\lvert u_s\rvert^2+\lvert u_r\rvert^2-\tilde{u}_1\tilde{u}_2\right]$
with $\tilde{u}_j=u_j-2iv_j$ for $j=1,2$, $u_r=u_0+i u_z$, and $u_s=u_x+i u_y$.}

The resulting form of Eqs.~\eqref{Eq::eq_spectre} - \eqref{Eq::val_p} for this specific model is given in Appendix \ref{Diag_M}. Equation \eqref{Eq::eq_spectre} can be solved numerically in all parameter regimes, whereas analytical solutions are possible only in limiting cases. The simplest form is obtained in the limit $d\to0$ where the scattering coefficients do not depend on energy. In that case, the spin-orbit coupling plays no role and one recovers the well-known result of a single spin-degenerate ABS with energy~\cite{beriSplittingAndreevLevels2008c}
\begin{align}
{\epsilon_{1\sigma}^{(0)}(\phi)}=\epsilon_0\equiv\sqrt{1-(T+S)\sin^2\frac{\phi}{2}},
\label{Eq::Energy_jonction_longueur_nulle}
\end{align}
where  $T=\lvert t\rvert^2$, $S=\lvert s\rvert^2$ and $\epsilon=E/\Delta$. It is interesting to note that, in this limit, only the total transmission $T+S$ matters \cite{beriSplittingAndreevLevels2008c,dimitrova20052d}.

\begin{figure}[ht!]
\centering
\includegraphics[width=0.9\linewidth]{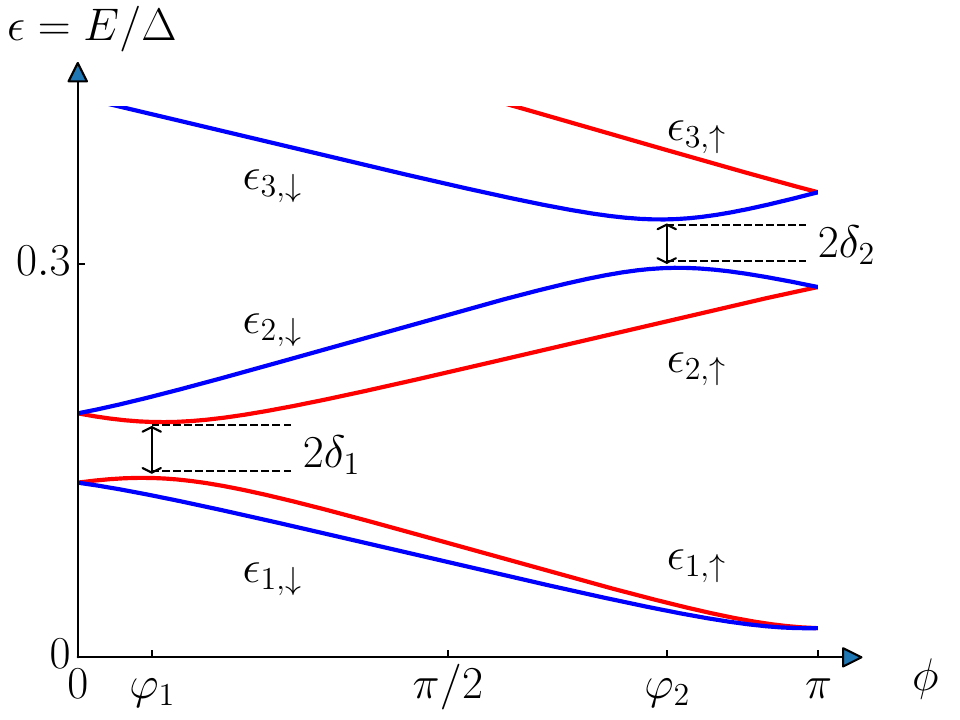}
\caption{\label{Fig::Figure_3}{Energy spectrum of ABS for a model with a single scattering center in the junction, {
shown up to an energy $\bar E\ll\Delta$. Here, {$T=0.95$, $S=0$, $\lambda_1=8$, $\lambda_2=10$, and $\tilde x_0=0.9$}}. Red lines correspond to spin-up states, while blue lines correspond to spin-down states. We denote the gap between doublet $m$ and $m+1$ at phase $\varphi_m$} {as $2\delta_m$.}}
\end{figure}

Spin-split Andreev levels are obtained, once one takes into account a finite length of the junction. This is illustrated in Fig.~\ref{Fig::Figure_3}, where the degeneracy of the ABS is lifted except for phases that are multiples of $\pi$, which effectively preserves TRS { and, hence, Kramers degeneracy}. To better understand the effect of the different parameters on this splitting, we {start by taking} the finite length of the junction into account perturbatively. Namely, we compute the corrections to $\epsilon_0$ (see. Eq.~\eqref{Eq::Energy_jonction_longueur_nulle})  to first order in {$\lambda_j=\Delta d/v_j$}, which yields
\begin{align}
&{\epsilon_{1\sigma}=\epsilon_0+\delta\epsilon_{\rm shift}+\sigma\delta\epsilon_{\rm split}} \label{Eq::eq_spectre_court}
\end{align}
with
\begin{align}
&\delta\epsilon_{\rm shift}=-\bar \lambda\epsilon_0\sqrt{1-\epsilon_0^2},\label{Eq::eq_spectre_court_shift}\\
&\delta\epsilon_{\rm split}=\frac{1}{2}|\delta\lambda|\sqrt{1-\epsilon_0^2}\sqrt{T+S\tilde x_0^2}\cos\frac\phi2,
\label{Eq::eq_spectre_court_split}
\end{align}
where $\bar\lambda=(\lambda_1+\lambda_2)/2$ and $\delta\lambda=\lambda_1-\lambda_2$. Here $\delta\epsilon_{\rm shift}$ describes a shift of both eigenvalues, whereas  $\delta\epsilon_{\rm split}$ describes the spin splitting. The splitting is proportional to $\delta\lambda$, highlighting the importance of the different Fermi velocities $v_1\neq v_2$. Furthermore, as $\sqrt{1-\epsilon_0^2}
\propto|\sin(\phi/2)|$, we recover the $\sin\phi$-dependence of the splitting predicted { perturbatively in SOC} in Refs.~\cite{chtchelkatchevAndreevQuantumDots2003b, beriSplittingAndreevLevels2008c}. Finally, we note that the splitting depends separately on the spin-conserving transmission $T$ and the spin-flip transmission $S$. At $R=0$ and $S\ll T$, up to small corrections $\propto S$, the result simplifies to 
\begin{equation}
\epsilon_{1\sigma}=\cos\frac\phi2-\frac12\left(\bar\lambda-{\frac\sigma2}\lvert\delta\lambda\rvert\right)\sin\phi. %{ +{\cal O}(S)}.
\label{Eq:spectrum-short}
\end{equation}
The above approximation is valid for phases not too close to zero. As can be observed in Fig.~\ref{Fig::Figure_3}, additional Andreev levels may appear in a finite length junction. {From Eq.~\eqref{Eq::eq_spectre}, setting $T=1$ and $\epsilon=1$, one can see that, for a short  junction, these additional states quickly join the continuum at $\phi=2\lambda_j$. Taking into account corrections up to second order in $\lambda_j$, one finds a crossing between the doublets $m=1$ and $m=2$ that takes place at $\varphi_{1}\approx|\delta\lambda|$}. A complementary view on the structure of Andreev levels in the vicinity of zero phase is discussed in Ref.~\cite{Yoko_singu}, see Fig.~9 therein.

In arbitrary length junctions, a simple expression for the low-energy spectrum, $\epsilon\ll1$,  can be obtained at $R=0$ and $S\ll T$, namely
\begin{equation}
   \epsilon_{m,\sigma}=\frac{f_m(\phi)}{2(1+\bar\lambda)+(-1)^{m}\sigma\lvert\delta\lambda\rvert}
   \label{Eq::energy_longue_jonction}
\end{equation}
where
$$f_m(\phi)=\begin{cases}m\pi-\phi,&m\enspace{\rm odd},\\(m-1)\pi+\phi, &m\enspace{\rm even}.\end{cases}$$
In the short junction limit, $\bar\lambda,\delta\lambda\ll1$, Eq.~\eqref{Eq::energy_longue_jonction} coincides with Eq.~\eqref{Eq:spectrum-short} { for $\phi$ near $\pi$ when the condition $\epsilon\ll 1$ is verified.}    
Time-reversal invariance at phases $\phi=0,\pi$ imposes level crossings at these phases. Namely,  { $\epsilon_{2m, \sigma}=\epsilon_{2m-1, -\sigma}$ at $\phi=0$, whereas $\epsilon_{2m, \sigma}=\epsilon_{2m+ 1, -\sigma}$  at $\phi=\pi$. } 

In the long junction limit, $\bar\lambda\gg1$, and assuming $\delta\lambda\ll\bar\lambda$, crossings between the same spin-states of doublets $m$ and $m+1$ occur at phases
\begin{align}
   \varphi_{m}=\begin{cases}
       \pi m\frac{\lvert\delta\lambda\rvert}{2\bar{\lambda}}, &m\enspace{\rm odd},\\
       \pi-\pi m\frac{\lvert\delta\lambda\rvert}{2\bar{\lambda}}, &m\enspace{\rm even}.
   \end{cases}\label{Eq::phi_crossing}
\end{align}
Note that for $m$ odd the crossing is between spin-up states whereas for $m$ even the crossing is between spin-down states.

These crossings are not protected by time-reversal invariance and are lifted at finite $R$ as we will see in the following. The resulting spectrum resembles the one shown in Fig.~\ref{Fig::Figure_3} at phases $\varphi_{m}<\phi<\varphi_{m+1}$  for $m$ odd and $\varphi_{m-1}<\phi<\varphi_{m}$ for $m$ even.

The energy gaps $\delta_m$, in units of $\Delta$, at the anti-crossings can be obtained by calculating the energy perturbatively in $R$ at the phases $\phi_c=\varphi_{m}$ with the help of Eq.~\eqref{Eq::eq_spectre}. We find
\begin{align}
   \delta_{m}=\frac{\sqrt R}{\bar\lambda}\begin{cases}
      \lvert \sin(\pi m\frac{\tilde x_0}{2})\rvert, &m\enspace{\rm odd},\\
       \\
       \lvert\cos(\pi m\frac{\tilde x_0}{2})\rvert, &m\enspace{\rm even}.
   \end{cases}\label{Eq::gap_R}
\end{align}
These gaps close at particular values of $\tilde x_0$, when the reflected and transmitted part of the wavefunction acquire the same phase through propagation in the normal region. 

The energies of the two states involved in the anti-crossing are then given as
\begin{align}
 {\epsilon^{>/<}_m=\frac{1}{2}\left(\epsilon^-_{m}+\epsilon^+_{m}\pm\sqrt{(\epsilon^-_{m}-\epsilon^+_{m})^2+4\delta_m^2}\right),}\label{Eq:eps-R}
\end{align}
where {$\epsilon^\pm_{m}$} corresponds to the energy level with positive/negative slope as a function of $\phi$. For $m$ odd,
\begin{align}
   \epsilon^\pm_{m}=
       \frac{\pi m\pm\phi}{2\bar\lambda}\left(1\mp\frac{\lvert\delta\lambda\rvert}{2\bar\lambda}\right),  
       \end{align}
whereas for $m$ even,
\begin{align}
   \epsilon^\pm_{m}=
       \frac{\pi (m\mp 1)\pm\phi}{2\bar\lambda}\left(1\pm\frac{\lvert\delta\lambda\rvert}{2\bar\lambda}\right).
       \end{align}
Note that $\epsilon^-_{m}-\epsilon^+_{m}=(\varphi_m-\phi)/\bar\lambda$ for all $m$. 

{The doublet $m=1$ requires special attention. At $T=1$, it crosses with the negative energy states at phase $\phi=\pi$, leading to a four-fold degeneracy at the Fermi level. Finite back-scattering opens up a gap (while preserving the two-fold degeneracy imposed by time-reversal symmetry). Using the same method as outlined above, we find that the positive energy states are shifted to $\delta_\pi=\sqrt{R}/\bar\lambda$.

\section{\label{Curr_op} Current Operator}

We now turn to the evaluation of {the matrix elements of the current operator}. Namely, we are interested in transitions between Andreev levels when a microwave drive is applied to the junction. {In particular, we will limit ourselves to the odd-parity sector, since we are interested in the spin-flip transitions between excited levels}. The microwave drive leads to a variation of the phase difference $\phi$ across the junction. If the variation $\delta\phi$ is small, we may linearize, {$H= H_{\rm BdG}+\delta\phi(\Phi_0/2\pi)\hat{J}$}, where $\hat{J}$ is the current operator given by
\begin{align}
\hat{J}=\frac{2\pi}{\Phi_0}\frac{\partial H_{\rm BdG}}{\partial\phi},
\label{Eq::def_operateur_courant}
\end{align}
and $\Phi_0=h/2e$ is the (superconducting) flux quantum~\cite{Trif2018,olivaresDynamicsQuasiparticleTrapping2014b}. {The coupling of the junction to the microwave drive is thus described by the current operator}, and its off-diagonal elements in the basis of Andreev levels determine which transitions can be induced. 

Using a gauge transformation {$H_{\rm BdG}\to  {\tilde H_{\rm BdG}}=e^{-i{{\phi }g(x)}\tau_z/2}H_{\rm BdG}e^{i{{\phi }g(x)}\tau_z/2}$, Eq.~\eqref{Eq::H_bdg} can be brought into the form
\begin{align}
\!\!\!\tilde {H}_{BdG}=H_e{\tau}_z+\frac{\phi}{2}\frac{\partial g(x)}{\partial x}\begin{pmatrix}v_1\sigma_z&0\\0&-v_2\sigma_z\end{pmatrix}\tau_0+\Delta(x){\tau}_x,
\label{Eq::def_H_BdG_tilde}
\end{align}
where $g(x)$ describes the phase profile along the $x$ direction with $g(d)=-g(0)=1/2$.} Therefore, the current operator may be written as
\begin{equation}
\hat J=\frac{\pi}{\Phi_0}\frac{\partial g(x)}{\partial x}\begin{pmatrix}v_1\sigma_z&0\\0&-v_2\sigma_z\end{pmatrix}\tau_0.
\end{equation}
The matrix elements of the current operator are given by
\begin{align}
{J}_{nn'}=\int  dx\;\Psi^\dagger_n(x)\hat J\Psi_{n'}(x),
\label{Eq::Current operator matrix_elements}
\end{align}
where $\Psi_n(x)$ is the wave function of the Andreev level $n$ {associated with the spectrum obtained in Sec.~\ref{Sec::Energy_spectrum}} (see Appendix \ref{Diag_M}) and $n=(m,\sigma)$ is a composite index. For the diagonal elements, the expression simplifies to $J_{nn}=e\partial_\phi E_n$ as expected from the Feynman-Hellmann theorem.

To evaluate the off-diagonal elements of the current operator, we need to know how the phase of the superconducting order parameter drops along the nanowire. {Under ac drive, determining the profile of {$g(x)$}, which in turn determines the electric field inside the nanowire, requires an involved self-consistent calculation that takes into account electron-electron interactions in the system, see, e.g., \cite{blanter1998}. A common simplifying assumption is that the charge inside a nanowire with low electron density is fully screened by nearby metallic gates, see, e.g., \cite{Safi2008,Gaury2014}. Then, the phase profile along the nanowire becomes frequency-independent and can be determined by an electrostatic calculation that involves the capacitance matrix between the leads and the gates.} Rather than evaluating the electrostatic profile along the nanowire, we will compute the elements of the current operator for the case when the entire phase drop happens at an arbitrary point $x'$, i.e.,
{\begin{align}
   &g(x)=\theta(x-x')-1/2,
\label{Eq::choix_chute_phase}
\end{align}
such that  $\partial g(x)/\partial x=\delta(x-x')$}. The off-diagonal elements of the current operator for different phase profiles can be obtained by appropriately averaging over $x'$. {In particular, in the case of a single metallic screening gate, it is expected that the phase drop takes place at {the interfaces between the nanowire and the leads} at $x'=0$ or $x'={d}$. In Figs.~\ref{Fig::figure_4} and \ref{Fig::Figure_6}-\ref{Fig::Figure_optimal}, {we show results for a phase drop at either one or the other interface}. More complicated phase profiles can arise in the case of several gates: in Fig. \ref{Fig::Figure_10}, we will assume that the phase drop arises at a single arbitrary position along the nanowire, while a linear drop of the phase was assumed in Ref.~\cite{haysCoherentManipulationAndreev2021}.} Let us note that the phase profile does not affect the value of the diagonal elements of the current operator due to the Feynman-Hellmann theorem mentioned earlier.

Using the wave functions given in Appendix \ref{Diag_M}, the current operator elements can be written in the following form
\begin{widetext}
\begin{align}
  {{J}_{nn'}}=e\sqrt{N_n N_{n'}}
       \begin{cases}
   \underset{k=1,2}{\sum}\left[
  f_{nn',k}^+\mathcal{A}_{(n)k}^*\mathcal{A}_{(n')k}
   -f_{nn',k+1}^+\alpha_n^*\alpha_{n'}({\cal S}_{e(n)}\mathcal{A}_{(n)})_k^*({\cal S}_{e(n')}\mathcal{A}_{(n')})_k
 \right], &\text{$0<x'<x_0$},\\
   \\
    - \underset{k=3,4}{\sum}\left[
  f_{nn',k-1}^-\mathcal{A}_{(n)k}^*\mathcal{A}_{(n')k}
   -f_{nn',k}^-\alpha_n^*\alpha_{n'}({\cal S}_{e(n)}\mathcal{A}_{(n)})_k^*({\cal S}_{e(n')}\mathcal{A}_{(n')})_k
 \right], &\text{$x_0<x'<d$},
   \end{cases}
   \label{Eq::eq_courant_1particule}
\end{align}
\end{widetext}
where $f_{nn',k}^\pm=e^{i\lambda_k(\epsilon_{n'}-\epsilon_n)({1\pm\tilde{x}'})/{2}}+\alpha_n \alpha_{n'}^* e^{-i\lambda_k (\epsilon_{n'}-\epsilon_n)({1\pm\tilde{x}'})/{2}}$
with $\tilde{x}'=2x'/d-1$, and we defined $\lambda_3=\lambda_1$, $\lambda_4=\lambda_2$. Furthermore, $N_n$ is the normalization coefficient of the $n$-th ABS, and $\mathcal{A}_{(n)k}$ the component $k$ of the eigenvector of $M(E)$ associated with the state $n$, see Appendix~\ref{Diag_M}.

It is important to note that Eq.~\eqref{Eq::eq_courant_1particule} can be simplified when certain components $\mathcal A_{(n)k}$ of the wavefunction are zero. In particular, this may lead to the absence of specific transitions. If $S=0$, the spin-up and spin-down states are decoupled and all spin-flip transitions are absent. { A similar decoupling into two independent blocks occurs when there is only spin-flip scattering ($T=0$)}. If $R=0$, the states with different {parity of $m$}, i.e., with positive and negative slope as a function of $\phi$, are decoupled and all transitions between a doublet with $m$ odd and a doublet with $m$ even are absent.

Equation \eqref{Eq::eq_courant_1particule} allows one to compute all the elements of the current operator for arbitrary parameters numerically. Before showing the results, let us discuss limiting cases, where analytical results are possible due to the above-mentioned simplifications.
Time-reversal symmetry relates states at phases $\phi$ and $2\pi-\phi$. In particular, $\mathcal T\Psi_{m\sigma}(2\pi-\phi)=\Psi_{m-\sigma}(\phi)$  with $\mathcal T=\tau_0\Theta$, where $\Theta$ is defined in Eq.~\eqref{Eq::TRS_operator}. Using $\mathcal T \hat J\mathcal T^{-1}=-\hat J$, it follows that $J_{m\sigma\to m'\sigma'}(2\pi-\phi)=-J_{m-\sigma\to m'-\sigma'}(\phi)$, i.e., we can restrict ourselves to computing the current operator matrix elements in the phase interval $\phi\in[0,\pi]$.

\subsection{{Spin-flip transitions without backscattering}\label{Sec::current_op_perfect_transmission}}

The expressions for the elements of the current operator simplify considerably when considering the case $R=0$ and treating $S\ll1$ perturbatively. As in that case states with a different {parity of $m$} are decoupled, spin-flip transitions within a given doublet are only possible in the phase interval $\varphi_m<\phi<\varphi_{m+1}$ for $m$ odd and $\varphi_{m-1}<\phi<\varphi_{m}$ for $m$ even. For inter-doublet spin-flip transitions, the phase interval is given as $\varphi_{m'}<\phi<\varphi_{m-1}$ for $m,m'$ odd and as $\varphi_{m'-1}<\phi<\varphi_{m}$ for $m,m'$ even.
Within that interval, we can compute the matrix elements $J_{m\downarrow\to m'\uparrow}$ using the eigenvectors given by Eq.~\eqref{Eq:ev_S}. { For $m$ odd, only the components ${\cal A}_{(n)k}$ with $k=3,4$ contribute, whereas, for $m$ even  only the components ${\cal A}_{(n)k}$ with $k=1,2$ contribute. Furthermore, only the product ${\cal A}^*_{(m\downarrow)k}{\cal A}_{(m'\uparrow)k-1}$  is of order 1 whereas ${\cal A}^*_{(m\downarrow)k}{\cal A}_{(m'\uparrow)k}$ is of order $\sqrt S$.}
Keeping only terms up to order $\sqrt{S}$ and assuming {$|\delta\lambda|\epsilon,~ \delta\epsilon\ll1$, where $\delta\epsilon=\epsilon_{m'\uparrow}-\epsilon_{m\downarrow}$, one obtains after a lengthy but straightforward calculation the matrix elements for spin-flip transitions between doublets $m$ and $m'$, where $m+m'$ even. Up to a global phase factor, they take the form}{ (see Appendix \ref{Appendix_C})}
\begin{align}
         \frac{\left\lvert J_{m\downarrow\to m'\uparrow}\right\rvert}{\sqrt{N_\uparrow N_\downarrow}}=&\sqrt{S}\frac e2\lvert\delta\lambda\delta\epsilon\rvert\label{Eq::courant_x_dep_general}\\
   &\times {\left\lvert\frac{  \eta(\tilde x_0) }2 \lvert\delta\lambda\rvert\bar\epsilon f_\pm^{(0)}+(-1)^m (1\mp \tilde x_0)f_\pm^{(1)}\right\rvert}, \nonumber
\end{align}
with
\begin{align}
   f_\pm^{(0)}=&\cos\left[\frac{\delta\epsilon}{2}\left(\frac{1}{\sqrt{1-\bar\epsilon^2}}+\bar\lambda (1\pm\tilde x')\right)\right],\\
   f_\pm^{(1)}=&(1\pm\tilde x')\sin\left[\frac{\delta\epsilon}{2}\left(\frac{1}{\sqrt{1-\bar\epsilon^2}}+\bar\lambda (1\pm\tilde x')\right)\right],
\end{align}
and  $ \eta(\tilde x_0) =(1-\tilde x_0^2)(1\mp \tilde x_0/3)$. Here the upper (lower) sign has to be used for $x'<x_0$  ($x'>x_0$). Furthermore, $\bar\epsilon=(\epsilon_{m'\uparrow}+\epsilon_{m\downarrow})/2$.

{According to Eq.~\eqref{Eq::courant_x_dep_general}, the spin-flip current operator matrix elements vanish when the barrier is at one of the interfaces, $|\tilde x_0|=1$. { As discussed in Appendix~\ref{double_barrier_model}, this feature is true beyond the specific model considered here: if scattering is only taking place at the interfaces, there are no spin-flip transitions}. For $|\tilde x_0|\neq1$, Eq.~\eqref{Eq::courant_x_dep_general}  yields a finite result that will be analyzed in more detail for the case of short and long junctions below.

\subsubsection{Short junction}

In the short junction limit, the only possible transition is the intra-doublet transition $1\downarrow\to1\uparrow$. Using $\delta\epsilon,\bar\lambda\ll1$, the current matrix element Eq.~\eqref{Eq::courant_x_dep_general} further simplifies to
\begin{align}
 \lvert J_{1\downarrow\to 1\uparrow}\rvert=&\sqrt{S}\frac {e}{4}\sqrt{N_\uparrow N_\downarrow}\delta\lambda^2\rvert\delta\epsilon\,\bar\epsilon\rvert\\
   &\times \left\lvert{  \eta(\tilde x_0) }- (1\mp \tilde x_0)(1\pm\tilde x')\right\rvert. \nonumber
\end{align}
Using Eq.~\eqref{Eq:spectrum-short} to obtain $\bar\epsilon,\delta\epsilon$ and $\sqrt{N_\uparrow N_\downarrow}\approx\Delta\sin(\phi/2)/2$ as well as the expression for $ \eta(\tilde x_0) $, one finds 
{\begin{align}
 \lvert J_{1\downarrow\to 1\uparrow}\rvert=&\sqrt{S}\frac {e\Delta}{32}|\delta\lambda|^3\sin^2\phi\label{eq:zero-short}\\
   &\times (1\mp\tilde x_0)\left\lvert \tilde x_0-\tilde x'-\frac13\tilde x_0(1\pm\tilde x_0)\right\rvert. \nonumber
\end{align}}
The characteristic scale for the magnitude of the current matrix element is {$J_{0}^{({\rm short})}=\sqrt Se\Delta|\delta\lambda|^3/32$}.

\begin{figure}[ht!]
   \centering
   \subfloat[]{\includegraphics[width=1\linewidth]{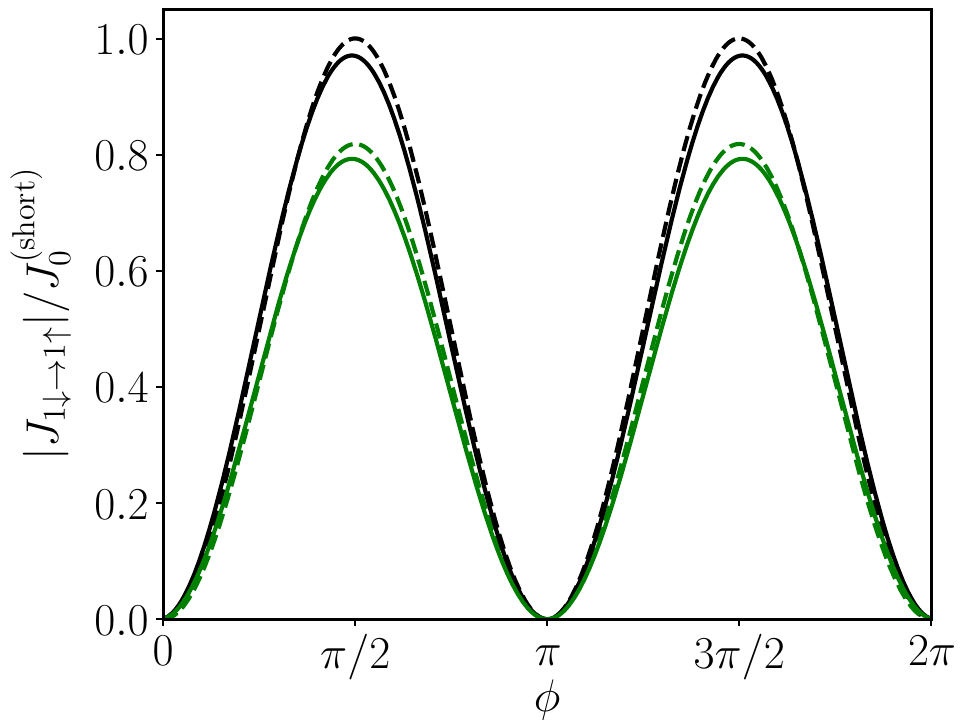} \label{Fig::figure_4_a}}\\
   \subfloat[]{\includegraphics[width=1\linewidth]{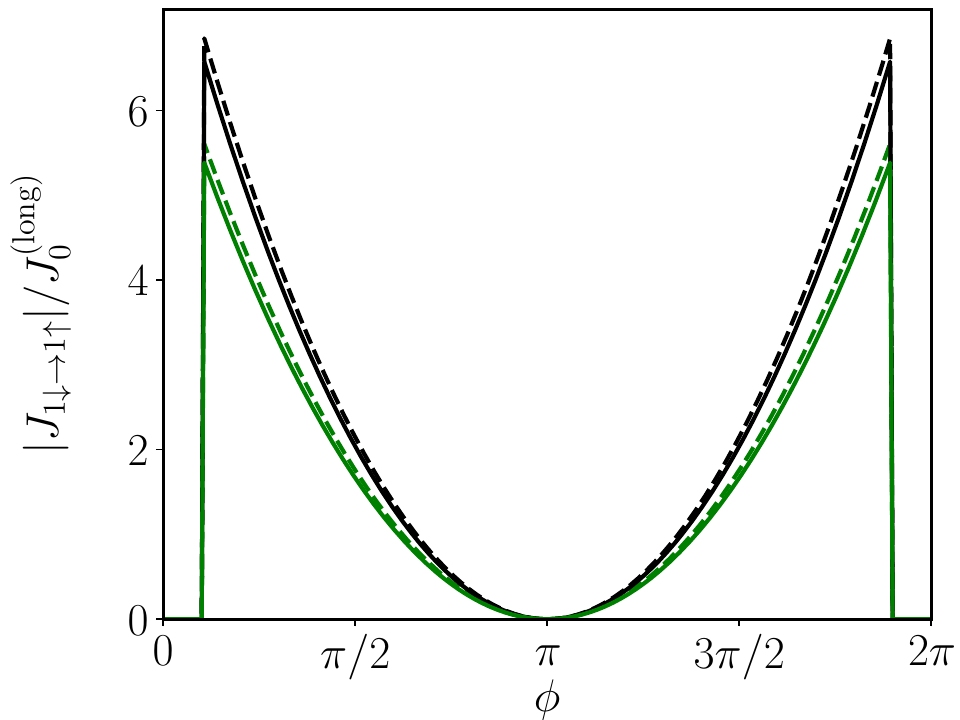}\label{Fig::figure_4_b}}
   \caption{{{Spin-flip current operator matrix elements within the lowest doublet in the absence of backscattering, $R=0$}. \ref{sub@Fig::figure_4_a} Short junction ($\lambda_1=0.02$ and $\lambda_2=0.01$) and \ref{sub@Fig::figure_4_b} long junction ($\lambda_1=20$ and $\lambda_2=16$). The parameters for both panels are $T=0.99$, $S=0.01$, and $\tilde x_0=0.3$. The current operator elements are normalized by $J_{0}^{({\rm short})}$ and $J_{0}^{({\rm long})}$ for the short and long junction, respectively.  Results for a phase drop at $x'=0$ (green) and $x'=d$ (black) are shown.
   Dashed lines correspond to the analytical results and full lines to the numerical results. As can be seen in panel \ref{sub@Fig::figure_4_b}, the matrix element sharply drops to zero at $\phi=\varphi_1$. In panel \ref{sub@Fig::figure_4_a}, the drop happens at a phase too close to zero to be visible.}}
   \label{Fig::figure_4}
   \end{figure}

\begin{figure}[ht!]
   \centering
   \includegraphics[width=1\linewidth]{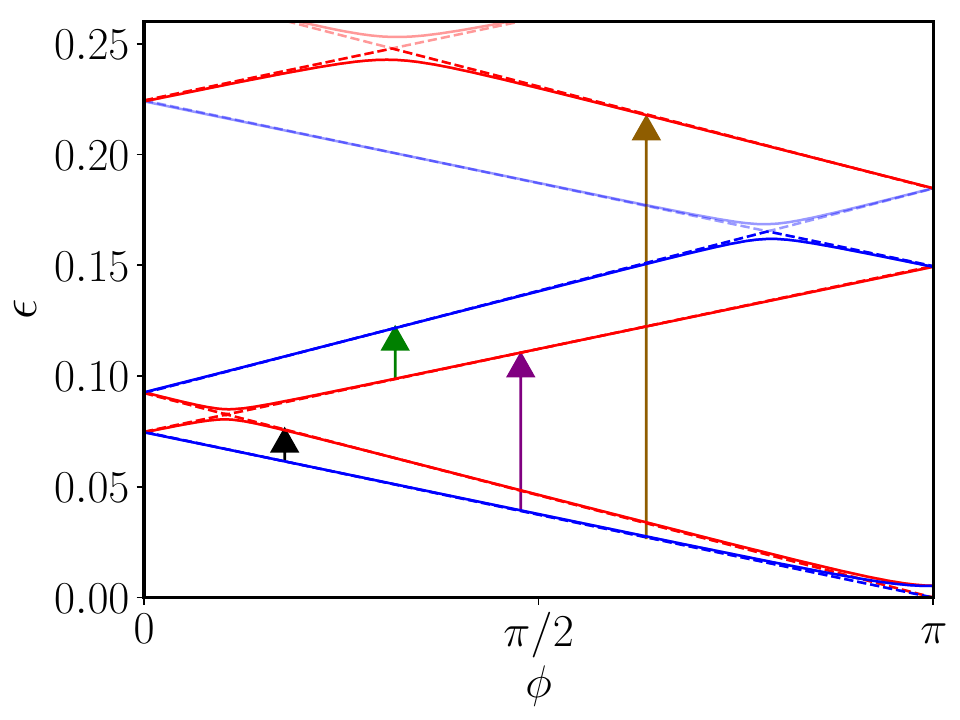}
   \caption{{Energy spectrum at perfect transmission ($R=0$, dashed lines) and with finite backscattering ($R=0.01$, full lines). The colored arrows indicate the transitions for which we calculate the matrix elements of the current operator in Figs.~\ref{Fig::Figure_6} and~\ref{Fig::Figure_8}. The parameters are $T=0.99-R$, $S=0.01$, $\tilde x_0=0.3$, $\lambda_1=20$, and $\lambda_2=16$.}}
   \label{Fig::Figure_5}
\end{figure}

\begin{figure}[ht!]
   \centering
   \includegraphics[width=1\linewidth]{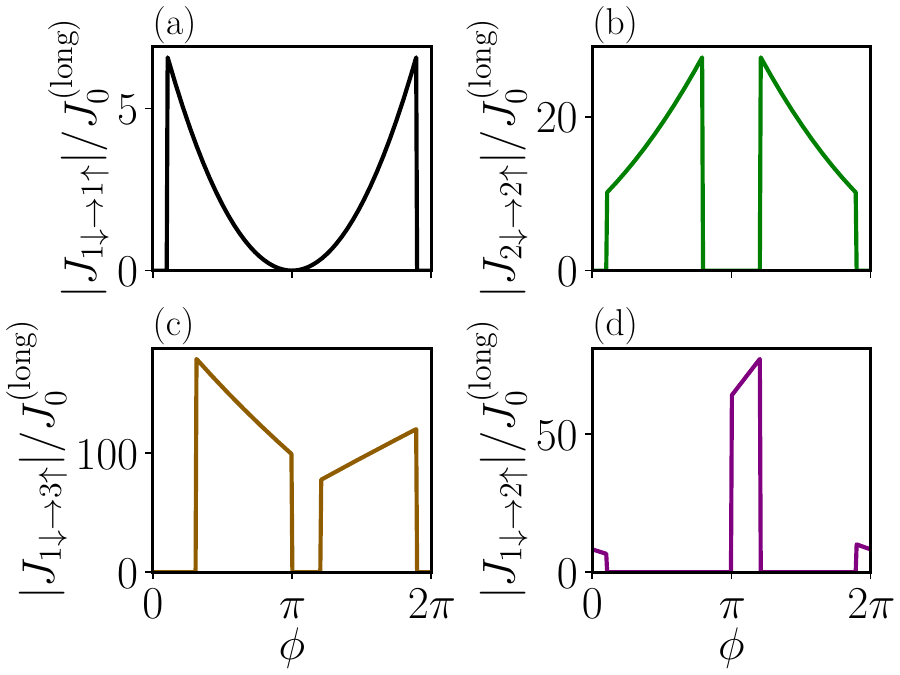}
   \caption{{Numerical results for the matrix elements of the current operator between opposite spin states at perfect transmission associated with the spectrum shown in Fig.~\ref{Fig::Figure_5}. Panels (a) and (b) show the intra-doublet matrix elements for the first and second doublet, respectively.  Panel (c) shows the inter-doublet  spin-flip matrix elements between the first and third doublet, having the same parity. Panel (d) shows the inter-doublet spin-flip matrix elements between the first and second doublet, having opposite parity. In all panels, we took $x'=d$. The abrupt drops to zero at phases $\varphi_m$ are due to the various crossings between states at perfect transmission.}}
   \label{Fig::Figure_6}
\end{figure}

\subsubsection{Long junction}

In the long junction limit, several doublets with energies $\epsilon\ll1$ exist and Eq.~\eqref{Eq::courant_x_dep_general} can be applied { in a large phase interval  {comprising $\pi/2$ up to the level crossings given by Eq.~\eqref{Eq::phi_crossing}}. Approximating Eq.~\eqref{Eq::energy_longue_jonction} as
$$\epsilon_{m,\sigma}\approx\frac{f_m(\phi)}{2\bar\lambda}\left(1-(-1)^m \sigma\frac{\lvert\delta\lambda\rvert}{2\bar\lambda}\right)$$
and $\sqrt{N_\uparrow N_\downarrow}\approx\Delta/(2\bar\lambda)$, 
one finds that $\delta\epsilon\approx|\delta\lambda|\bar\epsilon/\bar\lambda$  for the intra-doublet matrix elements, yielding
\begin{align}
 { \lvert J_{m\downarrow\to m\uparrow}\rvert}=&\sqrt{S}\frac {e\Delta}{32\bar\lambda}\left\lvert\frac{\delta\lambda}{\bar\lambda}\right\rvert^3f_m^2(\phi)\label{Eq::courant_x_dep_long}\\
   &\times \left\lvert  \eta(\tilde x_0)  +(-1)^m (1\mp \tilde x_0)(1\pm\tilde x')^2\right\rvert, \nonumber
\end{align}
and the phase interval is delimited by $\varphi_{m}<\phi<\varphi_{m-1}$ for $m$ odd and $\varphi_{m-1}<\phi<\varphi_{m}$ for $m$ even.} { In the long junction regime, the characteristic scale for the magnitude of the matrix elements of the current operator is $J_{0}^{({\rm long})}=\sqrt{S}eE_T\lvert\delta\lambda/\bar\lambda\rvert^3/32$ with the Thouless energy $E_T=\Delta/\bar\lambda$. As for the short junction, the amplitude is proportional to $\lvert\delta\lambda\rvert^3$. {Namely, it is suppressed as $\lvert\delta\lambda/\bar\lambda\rvert^3\ll1$. However, as expected in a long junction, the overall energy scale for the transition matrix elements is set by the Thouless energy rather than {the superconducting gap.}}}}

{By contrast, for the inter-doublet matrix elements, $\delta\epsilon\approx (m'-m)\pi/(2\bar\lambda)$. In that case, the two terms in the second line of Eq.~\eqref{Eq::courant_x_dep_general} behave differently. For $|\tilde x\pm1|\ll|\delta\lambda|/\bar\lambda$, the first term dominates and we obtain
{ \begin{align}
 \lvert J_{m\downarrow\to m+2n\uparrow}\rvert\approx&\sqrt{S}\frac{e\Delta}{16\bar\lambda}\left\lvert\frac{\delta\lambda}{\bar\lambda} \right\rvert^2\pi |n|  \eta(\tilde x_0) f_{m+n}(\phi) ,\label{Eq::courant_x_dep_long-inter1}
  \end{align}
whereas for other values of $\tilde x$, the second term dominates and the result reads}
\begin{align}
  \lvert J_{m\downarrow\to m+2n\uparrow}\rvert\approx&\sqrt{S}\frac{e\Delta}{4\bar\lambda}\left\lvert\frac{\delta\lambda}{\bar\lambda}\right\rvert \pi |n|\label{Eq::courant_x_dep_long-inter2}\\
   &\times (1\mp \tilde x_0)(1\pm\tilde x') \left \lvert \sin\left(n\frac \pi2(1\pm\tilde x')\right)\right\rvert. \nonumber
\end{align}
Let us first note that these matrix elements are larger than the intra-doublet matrix elements which have an additional suppression factor due to the small energy difference $\delta\epsilon\propto\delta\lambda$. Furthermore, their magnitude strongly depends on the phase profile. Namely, 
it is enhanced by a factor $|\bar\lambda/\delta\lambda\rvert\gg1$ when the phase drop is not at the interfaces.~\footnote{For $|n|>1$, the enhancement only holds when the phase drop occurs away from the positions $n(1\pm\tilde x')/2\in\mathbb{Z}$.}} {Figure~\ref{Fig::figure_4} shows the intra-doublet matrix elements of the current operator within the first doublet in the short and long junction regime, while Fig.~\ref{Fig::Figure_6} shows both intra-doublet and inter-doublet spin-flip matrix elements in the long junction regime. We show the matrix elements $\lvert J_{m\downarrow\to m'\uparrow}\rvert$ over the entire phase interval $\phi\in[0,2\pi]$. Using the relation $J_{m\sigma\to m'\sigma'}(2\pi-\phi)=-J_{m-\sigma\to m'-\sigma'}(\phi)$, the extended phase interval allows one to deduce the matrix elements  $\lvert J_{m\uparrow\to m'\downarrow}\rvert$ as well. The different transitions are indicated in Fig.~\ref{Fig::Figure_5}.}

\subsection{\label{Sec::effect_backscattering}Effect of finite backscattering}
{Finite backscattering couples states with different parity of $m$ and therefore renders the spin-flip current operator elements finite for all phases. This is particularly interesting in long junctions, where several doublets exist and anti-crossings take place at phases not too close to zero. To include the effect of backscattering on the current operator matrix elements in long junctions, we use the results of section \ref{Sec::Energy_spectrum} for the ABS at finite $R$. Namely, the wavefunctions corresponding to the energies given by Eq.~\eqref{Eq:eps-R} read}
\begin{align}
   \psi^{>}_m=&U_m\psi_{+\, m}+V_m\psi_{-\, m},\\
   \psi^{<}_m=&-V_m\psi_{+\, m}+U_m\psi_{-\, m},
\end{align}
{where $U_{m}=\Gamma_{m}/\sqrt{\left(\sqrt{\delta\varphi_m^2+\Gamma_m^2}-\delta\varphi_m\right)^2+\Gamma_m^2}$ and $V_{m}=\sqrt{1-U_{m}^2}$ with $\Gamma_m=2\bar\lambda\delta_m$ and $\delta\varphi_m=\phi-\varphi_m$. Thus, $U_m$ and $V_m$ vary around $\varphi_m$ on a typical scale set by $\Gamma_m$.}

\subsubsection{Modification of the intra- and inter-doublet spin-flip transitions in long junctions}

In this section, {we will focus on how the transitions we previously studied in section \ref{Sec::current_op_perfect_transmission} are modified due to finite backscattering.
{For intra- and inter-doublet matrix elements in the limit $\lvert 1\pm \tilde x'\rvert\ll\lvert\delta\lambda\rvert/\bar\lambda\ll 1$, Eq.~\eqref{Eq::courant_x_dep_general} simplifies to
\begin{align}
    { \lvert J_{m\downarrow\to m'\uparrow}\rvert}=&J_0\bar\lambda\left\lvert\epsilon_{m\downarrow}^2-\epsilon_{m'\uparrow}^2\right\rvert 
\end{align}
with $J_0=\sqrt{S}e\Delta(\delta\lambda/\bar
 \lambda)^2 \eta(\tilde x_0) /16$.}

 Finite back-scattering modifies this result in the vicinity of the anti-crossings on a scale $\Gamma_m$. If the different anti-crossings are well separated in phase on that scale,} we find the spin-flip matrix elements between doublets $m$ and $m'=m+2n$,
\begin{align}
\!\!\!\! \frac{ \lvert J_{m\downarrow\to m'\uparrow}\rvert}{ J_0 \bar\lambda}=\begin{cases}
       U_{m'} V_{m-1}\left\lvert\left(\epsilon^-_{m-1}\right)^2\!-\!\left(\epsilon^-_{m'}\right)^2\right\rvert&\!\!m\;{\rm odd},\!\!\\
       \\
       U_{m'-1} V_{m}\left\lvert\left(\epsilon^+_{m}\right)^2\!-\!\left(\epsilon^+_{m'-1}\right)^2\right\rvert&\!\!m\;{\rm even}.\!\!
   \end{cases}\label{Eq::J_spin_flip_finite_R_same_doublet_parity}
\end{align}
Thus, the main effect of finite backscattering is to smoothen the drop to zero over a width given by $\Gamma_m$, i.e., the typical scale of variation of $U_m$ and $V_m$. Figure \ref{Fig::Figure_8} shows numerical results for both intra-doublet and inter-doublet matrix elements $\lvert J_{m\downarrow\to m'\uparrow}\rvert$ over the entire phase interval $[0,\,2\pi]$. 
} 

{As discussed in Sec.~\ref{Sec::current_op_perfect_transmission}, the inter-doublet matrix elements are enhanced by a factor $\sim \bar\lambda/|\delta\lambda|$ for $|\tilde x\pm1|\gg|\delta\lambda|/\bar\lambda$. The smoothing due to finite backscattering involves the same factors $U_m$ and $V_m$, but starting form Eq.~\eqref{Eq::courant_x_dep_long-inter2} instead of Eq.~\eqref{Eq::courant_x_dep_long-inter1}.}

\subsubsection{Spin-flip matrix elements between opposite parity doublets}

In the absence of backscattering, spin-flip matrix elements between opposite parity doublets are possible only in a narrow phase interval around $0$ and $\pi$. Including backscattering renders them finite at all phases and  can be done the same way as in the previous section. { For a given transition, two anti-crossings are relevant, one close to zero and another one close to $\pi$.  The spin-flip matrix elements of the current operator between doublets $m$ and $m'=m+2n+1$ are given as
\begin{align}
\!\!\!\!  \frac{ \lvert J_{m\downarrow\to m'\uparrow}\rvert}{J_0\bar\lambda}= \enspace&
       \left\lvert U_{m'-1} U_{m-1}\left[\left(\epsilon^+_{m-1}\right)^2\!-\!\left(\epsilon^+_{m'-1}\right)^2\right]\right.\label{Eq::eq_50}\\
       &\left.+V_{m'-1} V_{m-1}\left[\left(\epsilon^-_{m-1}\right)^2\!-\!\left(\epsilon^-_{m'-1}\right)^2\right]\right\rvert\nonumber
\end{align}
for $m$ odd, and
\begin{align}
  \frac{ \lvert J_{m\downarrow\to m'\uparrow}\rvert}{J_0\bar\lambda}=&\left\lvert
       U_{m'} U_{m}\left[\left(\epsilon^-_{m}\right)^2-\left(\epsilon^-_{m'}\right)^2\right]\right.\label{Eq::eq_51}\\
       &\left.+V_{m} V_{m'}\left[\left(\epsilon^+_{m}\right)^2-\left(\epsilon^+_{m'}\right)^2\right]\right\rvert\nonumber,
\end{align}
for $m$ even. Here the first line {in Eq.~\eqref{Eq::eq_50} [in Eq.~\eqref{Eq::eq_51}]} is significant at phases $\phi\sim\varphi_{m-1} \enspace (\varphi_{m})$ while the second line is significant at phases $\phi\sim\varphi_{m'-1}  \enspace (\varphi_{m'})$ when $m$ is odd (even). As previously, the main effect of finite backscattering is to smoothen the drop to zero of the different matrix elements over a width $\Gamma_m$ around each crossing. An illustration of these matrix elements is shown in panel (d) of Fig.~\ref{Fig::Figure_8}.
\begin{figure}[ht!]
   \centering
   \includegraphics[width=1\linewidth]{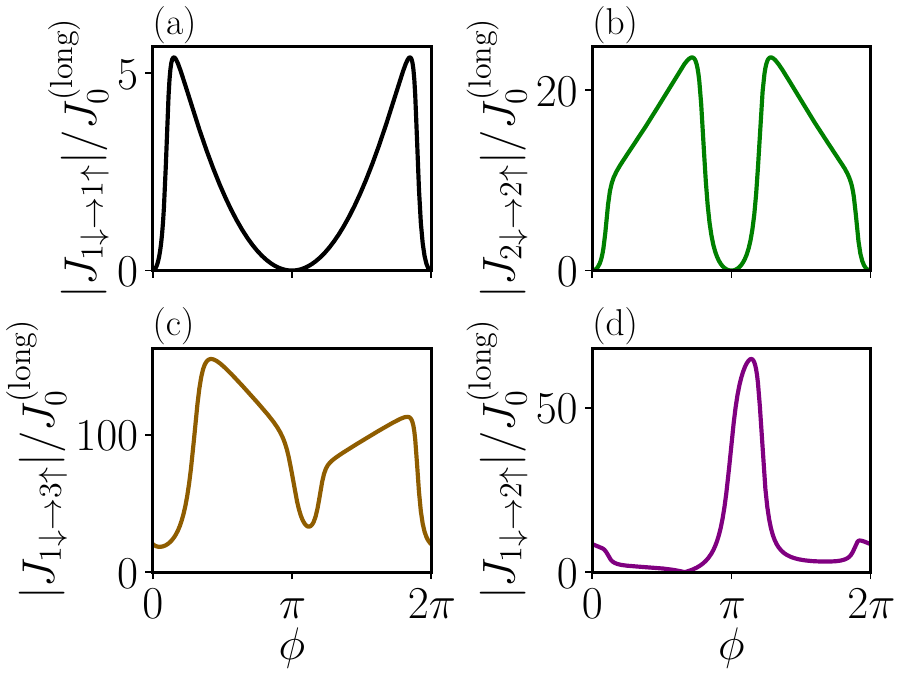}
   \caption{{Numerical results for the matrix elements of the current operator between opposite spin states at finite backscattering ($R=0.01$) associated with the spectrum shown in Fig.~\ref{Fig::Figure_5}.  As discussed in the main text, finite backscattering smoothens the sharp drops seen in Fig.~\ref{Fig::Figure_6}. In panel (c), one further sees that the matrix element no longer vanishes at phases close to $\pi$. This results from the avoided crossing between positive and negative energy states at $\varphi=\pi$ in the presence of finite backscattering.}}
\label{Fig::Figure_8}
\end{figure}

\subsection{Spin-conserving matrix elements}
Finally, we can look at the spin-conserving matrix elements of the current operator. {These matrix elements do not require spin-flip scattering. We will therefore start by calculating them at $R=S=0$. Then we will include backscattering as {in section \ref{Sec::effect_backscattering}.}

At $R=S=0$ only one component $\mathcal A_{(n),k}$ with $k=1,...,4$ is non zero, which simplifies Eq.~\eqref{Eq::eq_courant_1particule} significantly. As before, at $R=0$, only transitions between doublets with the same parity are allowed in a wide phase interval. Hence, the matrix elements between doublet $m$ and $m'=m+2n$ are given as
{\begin{align}
   \lvert J_{m\uparrow\rightarrow m'\uparrow}\rvert\approx&~ e E_T\left\lvert \cos\left[\frac{\pi n}{2}(1- \tilde x') \right] \right\rvert\\
   &\times\begin{cases}
       1+\mathrm{sign}(\phi-\varphi_m)\lvert\delta\lambda\rvert/(2\bar\lambda),&\!\!m\;{\rm odd},\!\!\\
       \\
       1-\mathrm{sign}(\phi-\varphi_{m-1})\lvert\delta\lambda\rvert/(2\bar\lambda),&\!\!m\;{\rm even}.\!\!\\
   \end{cases}\nonumber
\end{align}
For $\sigma=\,\downarrow$, one has to interchange $\mathrm{sign}(\phi-\varphi_m)$ and $\mathrm{sign}(\phi-\varphi_{m-1})$.}
For $m'=m+2n+1$, the same result holds, but in the complementary phase intervals where $\lvert J_{m\uparrow\rightarrow m+2n\uparrow}\rvert=0$.
As for the spin-flip matrix elements, the spin-conserving matrix elements sharply drop to zero at level crossings. Including backscattering smoothens these drops as discussed above for the spin-flip matrix elements. 

However, there is a particular case when $m'=m+1$. In that case, the two states involved in the transition cross at $\varphi_m$ in the absence of backscattering.  Backscattering mixes them and therefore enables transitions. One finds 
\begin{align}
   \lvert J_{m\sigma\to m+1\sigma}\rvert=2 e E_T\left\lvert U_m V_m \right\rvert.
\end{align}
This leads to a peak in the amplitude of the current operator matrix element at $\phi=\varphi_m$ as shown in Fig.~\ref{Fig::Figure_9}(a).
}

\begin{figure}[ht!]
   \centering
   \includegraphics[width=\linewidth]{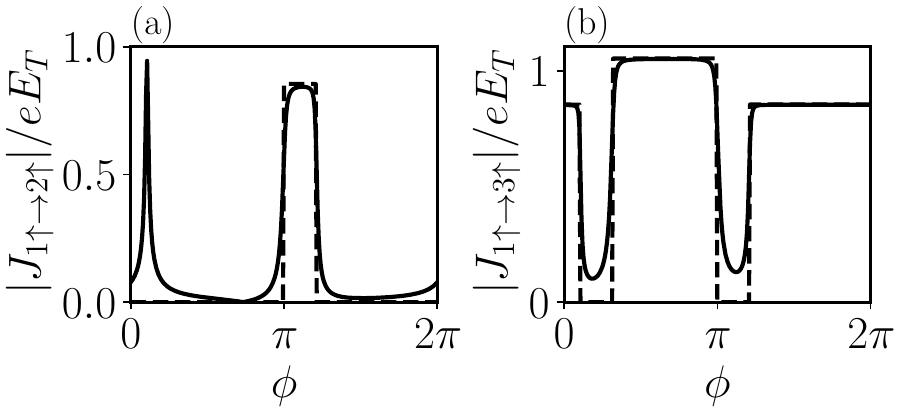}
   \caption{{Numerical results for the spin-conserving matrix elements of the current operator. Here $S=0$, $\tilde x_0=0.3$, $\tilde x'=1$, $\lambda_1=20$, and $\lambda_2=16$. The dashed and full lines correspond to perfect transmission $R=0$ and finite backscattering $R=0.001$, respectively. Panel (a) shows the matrix elements between spin-up states of the first and second doublet. The peak at $\phi=\varphi_1$ results from the mixing between the two states involved in the transition in the presence of finite backscattering. Panel (b) shows the matrix elements between spin-up states of the first and third doublet.}}
\label{Fig::Figure_9}
\end{figure}

\subsection{Numerical results}

{ Arbitrary length junctions and/or arbitrary values of the scattering parameters may be studied numerically using Eqs.~\eqref{Eq::eq_spectre} and \eqref{Eq::eq_courant_1particule}. In particular, we will be interested in the case when $S$ and $T$ are comparable and/or when $\delta\lambda\sim\bar\lambda$. A sample spectrum is shown in Fig.~\ref{Fig::Figure_optimal}(a) and the corresponding current operator matrix element for spin-flip transitions within the lowest doublet in Fig.~\ref{Fig::Figure_optimal}(b). The phase dependence is similar to the perturbative case with maxima at the avoided crossings. For the example shown, the scale for the magnitude of the current operator matrix elements is set by $J_0^{\rm (long)}=\sqrt{S}eE_T\lvert\delta\lambda/\bar\lambda\rvert^3/32$, where $eE_T$ is the relevant scale for the critical current of the junction. The smallness of the prefactor is due to numerical factors and does not contain a small parameter.

\begin{figure}[ht!]
   \centering
   \includegraphics[width=\linewidth]{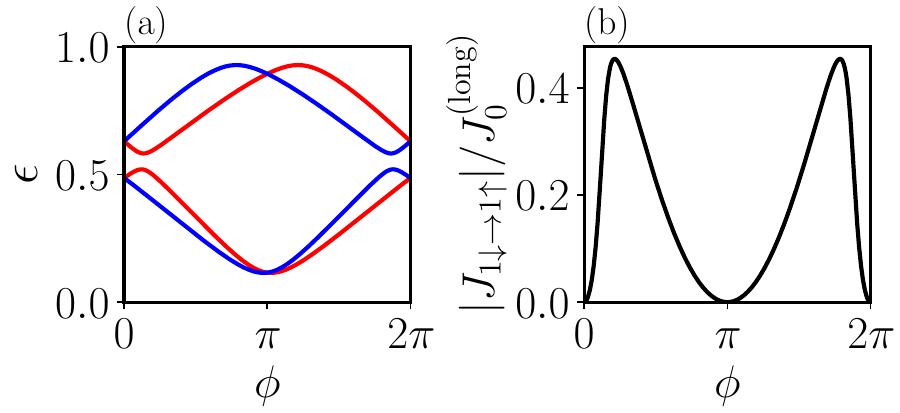}
       \caption{{Numerical results in the non-perturbative regime: (a) spectrum and (b) current operator matrix element for spin-flip transitions within the lowest doublet. Here $S=T=0.45$, $R=0.1$, $\lambda_1=2.3$, $ \lambda_2=1.3$, $\tilde x_0=0.3$, and $\tilde x'=1$. The phase dependence is similar to the perturbative case with maxima at the avoided crossings. The scale for the magnitude of the current operator matrix elements is set by $J_0^{\rm (long)}=\sqrt{S}eE_T\lvert\delta\lambda/\bar\lambda\rvert^3/32$.}}
\label{Fig::Figure_optimal}
\end{figure}

In Fig.~\ref{Fig::Figure_10}, we show the dependence of the magnitude of the current operator matrix element for intra-doublet transitions within the lowest doublet $m=1$ on various parameters. 
Panel (a) shows that the magnitude of the current operator matrix element is indeed maximal when $S$ and $T$ are comparable, whereas it vanishes when one of them is zero. Panel (b) shows the length dependence of the effect. As expected, intermediate length junctions are optimal. If the junction is too short, the effect of spin-orbit coupling is weak such that the magnitude of the spin-flip current operator matrix elements is suppressed. If the junction is too long, the overall energy scale for all the current operator matrix elements set by the Thouless energy is small and therefore suppresses the effect. Panel (c) shows the dependence of the magnitude of the current operator matrix element on the position of the scattering center. As mentioned earlier, spin-flip transitions are absent when scattering only occurs at the interfaces. Here we see that their amplitude is maximal when the scattering happens close to the center of the junction. Finally, panel (d) shows the variation of the current operator matrix element with the position of the phase drop.  The matrix element drops to zero for a particular value of $\tilde x$. { This can already be seen on the perturbative level; see Eqs.~\eqref{eq:zero-short} and \eqref{Eq::courant_x_dep_long}. In Appendix \ref{Appendix_C}, we show that the vanishing of the current operator matrix elements for particular values of $\tilde x$ generically happens also for other courant operator matrix elements, both spin-preserving and spin-flip.}

}

\begin{figure}[ht!]
\centering
\includegraphics[width=1\linewidth]{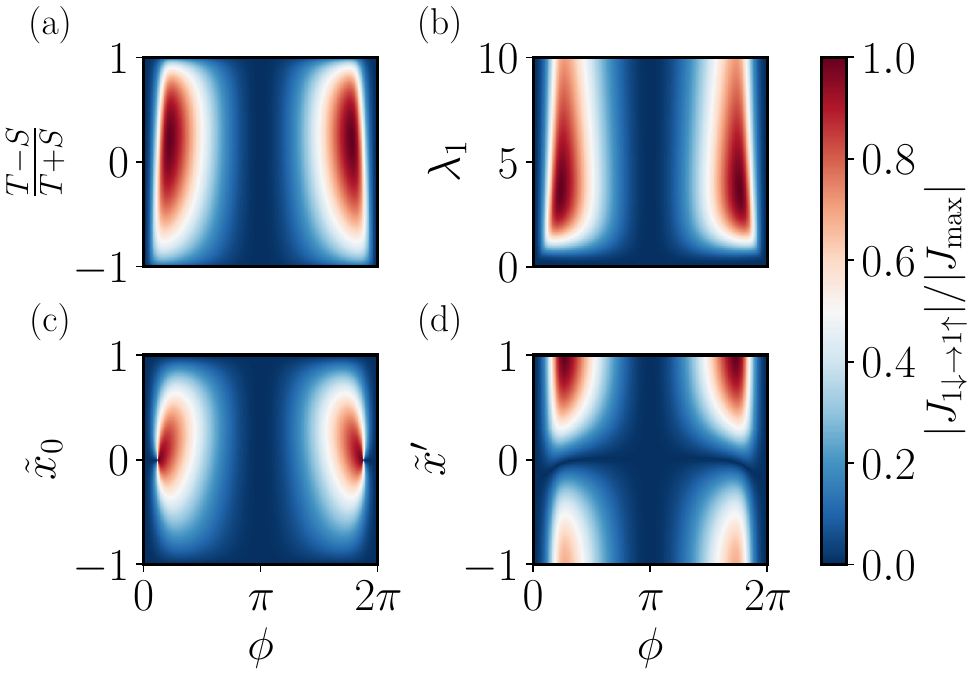}
\caption{ Dependence of the intra-doublet matrix elements of the current operator within the lowest doublet on different parameters. In all panels, {$R=0.1$} and $\lvert\delta\lambda\rvert/\bar\lambda=0.556$. In panel (a)-(c), the phase drop is set at $\tilde x'=1$. For panel (b)-(d), we have $T=S=0.45$. For panels (a), (c) and (d), we set $\lambda_1=2.3$. {For panels (a), (b), and (d), we set  $\tilde x_0=0.3$.}}
\label{Fig::Figure_10}
\end{figure}

\section{\label{Conclusion} Conclusion}

In this work, we estimated analytically and computed numerically the amplitude of the matrix elements of the current operator between two Andreev bound states forming an Andreev spin qubit in the odd-parity sector of a nanowire-based Josephson junction. These matrix elements characterize the coupling strength between the qubit and an external flux drive. In particular, they determine a variety of routinely measured observables, such as frequency shifts and resonance linewidths in microwave spectroscopy experiments, Rabi oscillations, and the  decoherence induced by quantum fluctuations in the electromagnetic environment of the qubit. 

We showed that generic scattering potentials yield non-vanishing matrix elements for all possible transitions, including the intra-doublet spin-flip transitions, in the absence of a magnetic field. The amplitude of the matrix element for intra-doublet spin-flip transitions is controlled by the spin-splitting of the spectrum and the presence of spin-flip scattering in the junction. Unless the system possesses additional symmetries, such scattering is generically present. Our findings indicate that the strong-coupling regime can be reached in a Josephson junction made with a nanowire of intermediate length (on the scale of the superconducting coherence length), provided that spin-orbit coupling (characterized by the relative asymmetry of the Fermi velocities in each of the pseudo-spin bands intercepting the Fermi level) is strong. Our results provide perspectives for the direct manipulation of an Andreev spin qubit with a single-tone drive, instead of the Raman protocol used in Ref.~\cite{haysCoherentManipulationAndreev2021}, which involves two tones and an auxiliary Andreev level.

In the nanowire-based Josephson junctions that we investigated in this work, we ignored the effect of Coulomb interaction. Thereby the Andreev spin qubit resides above the even ground state, and it requires a quasiparticle to \lq\lq poison\rq\rq\ the junction in order to be activated. Recent experiments with quantum dots subject to large Coulomb repulsion allowed stabilizing the doublet ground state in the odd sector, as well as resolving the spin splitting~\cite{bargerbos2022spectroscopy,PitaVidal2023,pitavidal2023strong}. Studying the current-operator matrix element for the operation of the associated spin qubit is an interesting direction for future investigation.

\begin{acknowledgements}
We thank {V. Fatemi} and H. Pothier for helpful discussions. Furthermore, we acknowledge funding from the French
Agence Nationale de la Recherche through Grant No.
ANR-21-CE30-0035 (TRIPRES) and ANR-23-CE47-0004 (FERBO).
\end{acknowledgements}

\appendix
\section{\label{Diag_M} Diagonalization of $M(E)$}

In this appendix, we provide the derivation of the eigenvalues and eigenvectors of $M(E)=r_A^*\mathcal{S}_h(E)r_A\mathcal{S}_e(E)$. From Eq.~\eqref{Eq::scattering_matrix}, we note 
\begin{align}
\rho(E)=r(E)\mathbb{1}_2,\qquad \tau(E)=\begin{pmatrix}
t(E) & s^*(E) \\
s(E) & -t^*(E)
\end{pmatrix}.
\end{align}
Hence, $M(E)$ takes the form
\begin{align}
M(E)&=e^{i\xi(E)-i\xi(-E)}\begin{pmatrix}
A & -B^\dagger \\
B &  D
\end{pmatrix}
\end{align}
with
\begin{align}
A&=\rho^\dagger(-E)\rho(E)+\tau^\dagger(-E)\tau(E)e^{-i\phi},\\
D&=\rho(-E)\rho^\dagger(E)+\tau(-E)\tau^\dagger(E)e^{i\phi},\\
B&= \rho(-E)\tau(E)-\tau(-E)\rho(E)e^{i\phi}.
\end{align}
From the unitarity of $M(E)$, we have $D=(B^\dagger)^{-1} A^\dagger B^\dagger$. Thus, $A$ and $D$ have a similar form which can be written as
\begin{align}
&A=
\begin{pmatrix}
\alpha_{A}+\beta_{A}e^{- i\phi} & -\delta_{A}^*e^{- i\phi} \\
\delta_{A} e^{- i \phi} & \alpha_{A}+\beta_{A}^*e^{- i\phi}
\end{pmatrix},\\
&D=
\begin{pmatrix}
\alpha_{D}+\beta_{D}e^{i\phi} & -\delta_{D}^*e^{ i\phi} \\
\delta_{D} e^{ i \phi} & \alpha_{D}+\beta_{D}^*e^{i\phi}
\end{pmatrix},
\end{align}
where
\begin{align}
\alpha_A&=\alpha_D^*=r_hr_e,\\
\beta_A&=t_ht_e+s_hs_e,\qquad\beta_D=t_h^*t_e^*+s_hs_e\label{Eq::elem_M_beta},\\
\delta_A&=s_h^*t_e-t_h^*s_e,\qquad\delta_D=s_h^*t_e^*-t_hs_e\label{Eq::elem_M_delta}.
\end{align}
With the scattering coefficients given in Eq.~\eqref{Eq::coef_scat_mat_s}, this yields
\begin{align}
\alpha_{A/D}&=R\,e^{\pm2i\bar\lambda\epsilon\tilde x_0},\\
\beta_{A/D}&=T\, e^{\pm i\delta \lambda\epsilon}+S\,e^{i\delta \lambda\epsilon\tilde x_0},\\
\delta_{A/D}&=\pm 2i\sqrt{TS}\,  e^{\frac{i}{2}\delta k_Fd(1\pm\tilde x_0)+i\varphi_s-i\varphi_t}\sin\frac{\delta \lambda\epsilon(1\mp\tilde x_0)}{2}.\label{Eq::elem_M_delta_1_barrier}
\end{align}
The matrices that diagonalize those blocks takes the form
\begin{align}
W_{A/D}=e^{-i\frac{\theta_{A/D}}{2}\sigma_z}e^{-i\frac{\gamma_{A/D}}{2}\sigma_y}
\end{align}
with
\begin{align}
&\!\!\!\theta_{A/D}=\mp\frac{\pi}{2}+\varphi_{{A/D}},\qquad\tan\gamma_{A/D}=\pm\frac{|\delta_{A/D}|}{\Im[\beta_{A/D}]},
\label{Eq::coef_diag_A_D}
\end{align}
where $\delta_{A/D}=|\delta_{A/D}|e^{i\varphi_{{A/D}}}$.

In the new basis, we will denote the four blocks as $\tilde{A}$, $\tilde{D}$, and $\tilde{B}$. One finds $\tilde D=\tilde A^\dagger$ with
\begin{align}
\!\!\!\tilde{A}= \;
& \alpha_{A}+\Re\left[\beta_{A}\right]e^{- i\phi} + i\sqrt{\Im^2[\beta_{A}]+|\delta_{A}|^2} e^{- i\phi}\,\sigma_z.
\label{eq::A16}
\end{align}
Note that, due to the square root in Eq.~\eqref{eq::A16} which results from the rotation of the block, { there is an ambiguity in the spin definition}. In particular, in the limit $T=1$, the square root simplifies to $\lvert\Im\left[\beta_A\right]\rvert$, which changes the way we label spin when $\Im\left[\beta_A\right]<0$.

Using unitarity, one concludes that  $\tilde A=\tilde B \tilde A \tilde B^{-1}$. Thus, the block $\tilde B$ must commute with $\tilde A$ and therefore be diagonal in the same basis, i.e.,
\begin{equation}
\tilde B=\begin{pmatrix}\tilde B_1&0\\0&\tilde B_2\end{pmatrix}.
\end{equation}
As a consequence, the diagonalization of the blocks $A$ and $D$ allows decomposing $M(E)$ into two $2\times 2$ independent blocks, which can be readily diagonalized to yield the eigenvalues in the form
$e^{2i\rho\chi_\sigma(\phi)+i\xi(E)-i\xi(-E)}$ with $\sigma,\rho=\pm1$ and
\begin{align}
\chi_\sigma(\phi)&=\arccos\sqrt{\frac{1+\tau\cos\left(\phi-\sigma\omega\right)+\Re\left[r_e r_h\right]}{2}},\\
\omega &={\rm{sign}(E)}\arccos\left(\frac{\Re\left[t_e t_h+s_e s_h\right]}{\tau}\right),\\
\tau&=\sqrt{(|t_e|^2+|s_e|^2)(|t_h|^2+|s_h|^2)},
\label{Eq::chi}
\end{align}
{which are the same as Eqs.~\eqref{Eq::chi_main_text}-\eqref{Eq::val_p}}. Here $t_{e/h}$, $s_{e/h}$ and $r_{e/h}$ are the scattering coefficients of electrons/holes related by $t_h(E)=t_e^*(-E)$, $s_h(E)=s_e^*(-E)$, and $r_h(E)=r_e^*(-E)$. 

The eigenvectors of $M(E)$ are given by
\begin{widetext}
{\begin{align}
       W=
       \begin{pmatrix}
        W_A & 0\\
        0 & W_D
    \end{pmatrix}\begin{pmatrix}
           \cos\frac{\gamma_{B1}}{2}e^{-i\frac{\theta_{B1}}{2}} & 0 & -\sin\frac{\gamma_{B1}}{2}e^{-i\frac{\theta_{B1}}{2}} & 0\\
           0 & \cos\frac{\gamma_{B2}}{2}e^{-i\frac{\theta_{B2}}{2}} & 0 & -\sin\frac{\gamma_{B2}}{2}e^{-i\frac{\theta_{B2}}{2}}\\
           \sin\frac{\gamma_{B1}}{2}e^{i\frac{\theta_{B1}}{2}} & 0 & \cos\frac{\gamma_{B1}}{2}e^{i\frac{\theta_{B1}}{2}} & 0 \\
           0 & \sin\frac{\gamma_{B2}}{2}e^{i\frac{\theta_{B2}}{2}} & 0 & \cos\frac{\gamma_{B2}}{2}e^{i\frac{\theta_{B2}}{2}}
       \end{pmatrix},
       \label{eigenvectors_M}
   \end{align}
}

\end{widetext}
where
\begin{align}
&\theta_{Bi}=\frac{\pi}{2}+\varphi_{Bi}, \qquad\tan\gamma_{B{i}}=-\frac{|\tilde{B}_{i}|}{\Im[\tilde{A}_{i}]}
\end{align}
with $\tilde B_i=|\tilde B_i|e^{i\varphi_{Bi}}$. Here the different columns correspond to different values of $(\rho,\sigma)$, namely the first column corresponds to the state $(-,+)$, the second column to  $(-,-)$, the third column to $(+,+)$, and the fourth column to $(+,-)$.

For the specific model with a single short-range scattering potential used here, the above equations can be simplified. In particular, Eqs.~\eqref{Eq::chi_main_text}-\eqref{Eq::val_p} can be reduced to 
\begin{align}
   &\chi_\sigma(\epsilon,\phi)=\arccos\sqrt{\tau\cos^2\frac{\phi-\sigma\omega}{2}+R\cos^2(\bar\lambda\epsilon\tilde x_0)},\\
   &\omega={\rm sign}(\epsilon)\arccos\left[\frac{T\cos(\delta\lambda\epsilon)+S\cos(\delta\lambda\epsilon\tilde x_0)}{\tau}\right],\\
   &\tau = T+S,
\end{align}
yielding the energy spectrum
\begin{eqnarray}
   \bar\lambda\epsilon+\rho\chi_\sigma(\epsilon,\phi)-\arccos\epsilon-q\pi=0.
\end{eqnarray}
The coefficients of $W$ on the other hand are given as
\begin{align}
   &\tan\gamma_{A/D}=\frac{2\sqrt{S T}\sin\left( \delta\lambda\epsilon(1\mp\tilde x_0)/2 \right)}{T\sin(\delta\lambda\epsilon)\pm S\sin(\delta\lambda\epsilon\tilde x_0)},\\
   &\tan\gamma_{B1/2}=\pm\frac{2\sqrt{R \tau}\sin\left( \frac{\phi\mp\omega}{2}+\bar\lambda\epsilon\tilde x_0\right)}{\tau\sin(\phi\mp\omega)-R\sin(2\bar\lambda\epsilon\tilde x_0)},\\
      &\theta_{A/D}=\delta k_F(1\pm \tilde x_0)d/2+\varphi_s\pm\varphi_t,\\
   &\theta_{Bi}(\phi)=\bar k_F \tilde x_0 d+\frac \phi2+\varphi_r+\frac{\pi}{2}\left(1+(-1)^i\right).
\end{align}
The eigenvectors of $M(E)$ allow one to obtain the wavefunctions of the ABS. Namely, the eigenvectors of $M(E)$ give the amplitudes of incoming electron states at the interfaces with the superconductors. To obtain the full wavefunction, we can construct the outgoing and hole amplitudes with the help of the normal and Andreev scattering matrices: 
\begin{align}
&\psi_{\rm out}^e=\mathcal{S}_e(E)\psi_{\rm in}^e,\\
&\psi_{\rm in}^h=\alpha(E)r_A(\phi)\mathcal{S}_e(E)\psi_{\rm in}^e,\\
&\psi_{\rm out}^h=\alpha^*(E)r_A(\phi)\psi_{\rm in}^e.
\end{align}
Then, using continuity, the wavefunctions in the nanowire and the superconductors can be computed. {Note that this result is independent of the scattering model, the only requirement is that it respect TRS}. In the basis in which the BdG-Hamiltonain in  Sec.~\ref{Sec::Model} is given, the wavefunctions for the case of a single scatterer at position $x_0$ take the following form in different regions:
\begin{itemize}
\item in the normal region of the nanowire to the left of the barrier, $0<x<x_0$:
\begin{align}
&\frac{\Psi_n(x)}{\sqrt{N_n}}=
\begin{pmatrix}
\mathcal{A}_{(n)1}e^{ik_1^e x}/\sqrt{v_1}\\
(\mathcal{S}_{e(n)}\mathcal{A}_{(n)})_2e^{-ik_1^e x}/\sqrt{v_1}\\
(\mathcal{S}_{e(n)}\mathcal{A}_{(n)})_1e^{-ik_2^e x}/\sqrt{v_2}\\
\mathcal{A}_{(n)2}e^{ik_2^e x}/\sqrt{v_2}\\
\alpha_n^*\mathcal{A}_{(n)1}e^{ik_1^h x+i\phi/2}/\sqrt{v_1}\\
\alpha_n(\mathcal{S}_{e(n)}\mathcal{A}_{(n)})_2e^{-ik_1^h x+i\phi/2}/\sqrt{v_1}\\
\alpha_n(\mathcal{S}_{e(n)}\mathcal{A}_{(n)})_1e^{-ik_2^h x+i\phi/2}/\sqrt{v_2}\\
\alpha_n^*\mathcal{A}_{(n)2}e^{ik_2^h x+i\phi/2}/\sqrt{v_2}
\end{pmatrix}, 
\end{align}
\item in the normal region of the nanowire to the right of the barrier, $x_0<x<d$:
\begin{align}
\frac{\Psi_n(x)}{\sqrt{N_n}}=
\begin{pmatrix}
(\mathcal{S}_{e(n)}\mathcal{A}_{(n)})_3e^{ik_1^e (x-d)}/\sqrt{v_1}\\
\mathcal{A}_{(n)4}e^{-ik_1^e (x-d)}/\sqrt{v_1}\\
\mathcal{A}_{(n)3}e^{-ik_2^e (x-d)}/\sqrt{v_2}\\
(\mathcal{S}_{e(n)}\mathcal{A}_{(n)})_4e^{ik_2^e (x-d)}/\sqrt{v_2}\\
\alpha_n(\mathcal{S}_{e(n)}\mathcal{A}_{(n)})_3e^{ik_1^h (x-d)-i\phi/2}/\sqrt{v_1}\\
\alpha_n^*\mathcal{A}_{(n)4}e^{-ik_1^h (x-d)-i\phi/2}/\sqrt{v_1}\\
\alpha_n^*\mathcal{A}_{(n)3}e^{-ik_2^h (x-d)-i\phi/2}/\sqrt{v_2}\\
\alpha_n(\mathcal{S}_{e(n)}\mathcal{A}_{(n)})_4e^{ik_2^h (x-d)-i\phi/2}/\sqrt{v_2}\\
\end{pmatrix},
\end{align}
{\item in the left superconductor, $x<0$:
\begin{align}
&\frac{\Psi_n(x)}{\sqrt{N_n}}=
\begin{pmatrix}
\mathcal{A}_{(n)1}e^{\kappa_{1n}x}/\sqrt{v_1}\\
(\mathcal{S}_{e(n)}\mathcal{A}_{(n)})_2e^{\kappa_{1n}x}/\sqrt{v_1}\\
(\mathcal{S}_{e(n)}\mathcal{A}_{(n)})_1e^{\kappa_{2n}x}/\sqrt{v_2}\\
\mathcal{A}_{(n)2}e^{\kappa_{2n}x}/\sqrt{v_2}\\
\alpha_n^*\mathcal{A}_{(n)1}e^{i\phi/2+\kappa_{1n}x}/\sqrt{v_1}\\
\alpha_n(\mathcal{S}_{e(n)}\mathcal{A}_{(n)})_2e^{i\phi/2+\kappa_{1n}x}/\sqrt{v_1}\\
\alpha_n(\mathcal{S}_{e(n)}\mathcal{A}_{(n)})_1e^{i\phi/2+\kappa_{2n}x}/\sqrt{v_2}\\
\alpha_n^*\mathcal{A}_{(n)2}e^{i\phi/2+\kappa_{2n}x}/\sqrt{v_2}
\end{pmatrix}, 
\end{align}
\item in the right superconductor, $x>d$:
\begin{align}
\frac{\Psi_n(x)}{\sqrt{N_n}}=
\begin{pmatrix}
(\mathcal{S}_{e(n)}\mathcal{A}_{(n)})_3e^{-\kappa_{1n}(x-d)}/\sqrt{v_1}\\
\mathcal{A}_{(n)4}e^{-\kappa_{1n}(x-d)}/\sqrt{v_1}\\
\mathcal{A}_{(n)3}e^{-\kappa_{2n}(x-d)}/\sqrt{v_2}\\
(\mathcal{S}_{e(n)}\mathcal{A}_{(n)})_4e^{-\kappa_{2n}(x-d)}/\sqrt{v_2}\\
\alpha_n(\mathcal{S}_{e(n)}\mathcal{A}_{(n)})_3e^{-i\phi/2-\kappa_{1n}(x-d)}/\sqrt{v_1}\\
\alpha_n^*\mathcal{A}_{(n)4}e^{-i\phi/2-\kappa_{1n}(x-d)}/\sqrt{v_1}\\
\alpha_n^*\mathcal{A}_{(n)3}e^{-i\phi/2-\kappa_{2n}(x-d)}/\sqrt{v_2}\\
\alpha_n(\mathcal{S}_{e(n)}\mathcal{A}_{(n)})_4e^{-i\phi/2-\kappa_{2n}(x-d)}/\sqrt{v_2}\\
\end{pmatrix}.
\end{align}}
\end{itemize}
{Here we defined $\kappa_{jn}=({\Delta}/{v_j})\sqrt{1-\epsilon_n^2}$}. In general, the subscript $n$ or $(n)$ indicates that a quantity is evaluated for a state with energy $\epsilon_n$, where $n=(m,\sigma)$ is a composite index.
The coefficients $\mathcal{A}_{(n)k}$ with $k=1\dots 4$ are the components of the eigenvector of $M(E)$ of state $n$. (Depending on the values of $(m,\sigma)$ the corresponding columns of the matrix $W$ have to be used.)

Some limiting cases will be useful. At $R=S=0$, the matrix $W$ reduces to
{
\begin{align}
   {W_{0}}=
   \begin{pmatrix}
       e^{-i\frac{{\theta_A}+\theta_{B1}}{2}}&0&0&0\\
       0&-ie^{i\frac{{\theta_A}-\theta_{B1}}{2}}&0&0\\
       0&0&e^{-i\frac{{\theta_D}-\theta_{B1}}{2}}&0\\
       0&0&0&ie^{i\frac{{\theta_D}+\theta_{B1}}{2}},
   \end{pmatrix}.
\end{align}
Introducing $R,S\ll1$ perturbatively yields {$W\approx W_0w_1$ with}
\begin{align}
\!\!\!\! {w_1}=\begin{pmatrix}1&i\sqrt{S}\,d_{s-}&\sqrt{R}\,d_{r-}&0\\
   i\sqrt{S}d_{s-}&1&0&\sqrt{R}\,d_{r+}\\
   -\sqrt{R}\,d_{r-}&0&1&-i\sqrt{S}\,d_{s+}\\
   0&-\sqrt{R}\,d_{r+}&-i\sqrt{S}\,d_{s+}&1\end{pmatrix}\!,\label{Eq:ev_S}
\end{align}
with $d_{s\pm}=\frac12\tan\gamma_{D/A}$ and $d_{r\pm}=\mp\frac12\tan\gamma_{B2/1}$. Using $\omega\approx\delta\lambda\epsilon$ and $\gamma_{A/D},\gamma_{B1/2}\ll 1$}, the corresponding expressions simplify to 
\begin{align}
   d_{s\pm}=&\frac{\sin\left[\frac{\delta\lambda\epsilon}2(1\pm\tilde x_0)\right]}{\sin(\delta\lambda\epsilon)},\\
d_{r\pm}=&\frac{\sin\left[\frac\phi2+(\bar\lambda\tilde x_0\pm\frac{\delta\lambda}2)\epsilon\right]}{\sin(\phi\pm\delta\lambda\epsilon){-}R\sin(2\bar \lambda \epsilon\tilde x_0)}.
\end{align}

\section{\label{double_barrier_model} Particular case of barriers at the superconductor interfaces}
From Eq.~\eqref{Eq::courant_x_dep_general}, we see that for a one scattering center model, the intra-doublet element vanishes for $\lvert\tilde x_0\rvert=1$. In this case, we see in Eqs.~\eqref{t} and~\eqref{s} that $s(E)$ and $t(E)$ have the same energy-dependent phase. As a consequence, the spectrum depends  only on the combination $T+S$, and the problem becomes analogous to having only one type of transmission. This can directly be seen by diagonalizing the transmission block of the scattering matrices. For the particular case of $\tilde x_0=1$, Eq.~\eqref{Eq::coef_diag_A_D} yields
\begin{align}
&\cos\gamma_{A,D}=\frac{T\pm S}{T+S}\\
&\sin\gamma_A=0,\quad \sin\gamma_D=\frac{2\sqrt{S T}}{T+S}
\label{Eq::coef_x0=1}
\end{align}
Those coefficients does not depend on the energy and 
\begin{align}
W_D^\dagger \tau(\pm E)W_A=\sqrt{T+S}e^{\pm\frac{1}{2}i\delta\lambda\epsilon\sigma_z}\sigma_z
\end{align}
As $W_D$ and $W_A$ become energy independent, Eq.~\eqref{Eq::eq_courant_1particule}. yields a vanishing result. 

For a model with a barrier at each interface, the results are similar. In that case, the scattering coefficients are given as
\begin{widetext}
\begin{align}
r(E)&=\frac{i}{|K(E)|}\left[r_L e^{-i\theta_{R}-i\bar k d}-r_R e^{i\theta_{L}+i\bar k d}\right],\\
t(E)&=\frac{i}{|K(E)|}\left[t_L t_R e^{\frac12i\delta k d}+s_L s_R^* e^{-\frac12i\delta k d}\right],\\
s(E)&=\frac{i}{|K(E)|}\left[t_L s_R e^{\frac{1}{2}i\delta k d}-s_L t_R^* e^{-\frac{1}{2}i\delta k d}\right],\\
\xi(E)&=\theta_{L}+\theta_{R}+\bar k d-\zeta(E)-\frac{\pi}{2},\\
\zeta(E)&=\arctan\left(-\frac{|r_L||r_R|\sin\left(\varphi_{\rm{tot}}+2\bar\lambda\epsilon\right)}{1-|r_L||r_R|\cos\left(\varphi_{\rm{tot}}+2\bar\lambda\epsilon\right)}\right),\\
|K(E)|^2&=\left(1-|r_L||r_R|\right)^2+4\lvert r_L\rvert\lvert r_R\rvert\sin^2\left(\frac{\varphi_{\rm{tot}}+2\bar\lambda\epsilon}{2}\right),\\
\varphi_{\rm{tot}}&=\theta_{L}+\theta_{R}+\varphi_{r_R}-\varphi_{r_L}+(k_{F1}+k_{F2})d.
\end{align}
\end{widetext}
Here $\theta_{{L/R}}$ are arbitrary global phases of the left/right barrier and $\varphi_{r_{R/L}}$ the phases of $r_{R/L}$. Again, $s(E)$ and $t(E)$ have the same energy dependency and as a consequence the matrices $W_A$ and $W_D$ are energy independent:
\begin{align}
\cos\gamma_{A/D}&=\frac{T_{L/R}-S_{L/R}}{T_{L/R}+S_{L/R}},\\
\sin\gamma_{A/D}&=\frac{2\sqrt{T_{L/R} S_{L/R}}}{T_{L/R}+S_{L/R}},
\end{align}
while
\begin{align}
   \theta_{A}=\varphi_{t_L}+\varphi_{s_L},\quad\theta_D=\varphi_{s_R}-\varphi_{t_R}.
\end{align}
As before, this yields zero spin-flip matrix elements when plugged into Eq.~\eqref{Eq::eq_courant_1particule}.

\section{\label{Appendix_C}Global phase of the current operator matrix elements}

{ Here we show that the global phase of the matrix elements of the current operator obtained from Eq.~\eqref{Eq::eq_courant_1particule} does not depend on $\tilde x'$.

The matrix $W$ containing the eigenvectors can be cast in the form $W=D_\varphi(\varphi_2)r_A(-\phi/2)\tilde W$, where 
\begin{widetext}
\begin{eqnarray}
   \tilde W=\begin{pmatrix}
   \cos\frac{\gamma_A}{2}\cos\frac{\gamma_{B1}}{2} & i\sin\frac{\gamma_A}{2}\cos\frac{\gamma_{B2}}{2} & -\cos\frac{\gamma_A}{2}\sin\frac{\gamma_{B1}}{2} & -i\sin\frac{\gamma_A}{2}\sin\frac{\gamma_{B2}}{2} \\
   \sin\frac{\gamma_A}{2}\cos\frac{\gamma_{B1}}{2} & -i\cos\frac{\gamma_A}{2}\cos\frac{\gamma_{B2}}{2} & -\sin\frac{\gamma_A}{2}\sin\frac{\gamma_{B1}}{2} & i\cos\frac{\gamma_A}{2}\sin\frac{\gamma_{B2}}{2} \\
   \cos\frac{\gamma_D}{2}\sin\frac{\gamma_{B1}}{2} & -i\sin\frac{\gamma_D}{2}\sin\frac{\gamma_{B2}}{2} & \cos\frac{\gamma_D}{2}\cos\frac{\gamma_{B1}}{2} & -i\sin\frac{\gamma_D}{2}\cos\frac{\gamma_{B2}}{2} \\
   \sin\frac{\gamma_D}{2}\sin\frac{\gamma_{B1}}{2} & i\cos\frac{\gamma_D}{2}\sin\frac{\gamma_{B2}}{2} & \sin\frac{\gamma_D}{2}\cos\frac{\gamma_{B1}}{2} & i\cos\frac{\gamma_D}{2}\cos\frac{\gamma_{B2}}{2}
   \end{pmatrix},
\end{eqnarray}   
\end{widetext}
and $D_\varphi(\varphi_2)$ is a diagonal-matrix containing energy-independent phases,
\begin{eqnarray}
D_\varphi(\varphi_2)=\exp\left[-\frac i2\left(\varphi_1\sigma_z+\varphi_2\tau_z+\varphi_3\sigma_z\tau_z\right)\right]
\end{eqnarray}
with $\varphi_1=\delta k_F d/2+\varphi_s$, $\varphi_2=\bar k_F d \tilde x_0+\varphi_r$ and $\varphi_3=\delta k_F d\tilde x_0/2 +\varphi_t$.
Thus, the phase of ${\cal A}_{(n)k}$ does not depend on the energy such that ${\cal A}_{(n)k}^*{\cal A}_{(n')k}$ is real.

Furthermore, we may use that the matrix 
\begin{widetext}
\begin{eqnarray}
S_e(E)=e^{i\theta+i\bar k(E) d}\begin{pmatrix}
   r e^{i\bar k(E) d\tilde x_0} & 0 & -t^*e^{-i\frac{\delta k(E)}{2}d} & -s^*e^{-i\frac{\delta k(E)}{2}d\tilde x_0}\\
   0 & r e^{i\bar k(E) d\tilde x_0} & -s e^{i\frac{\delta k(E)}{2}d\tilde x_0} & t e^{i\frac{\delta k(E)}{2}d}\\
   t e^{i\frac{\delta k(E)}{2}d} & s^*e^{-i\frac{\delta k(E)}{2}d\tilde x_0} & r^* e^{-i\bar k(E) d\tilde x_0} & 0 \\
   s e^{i\frac{\delta k(E)}{2}d\tilde x_0} & -t^* e^{-i\frac{\delta k(E)}{2}d} & 0 & r^* e^{-i\bar k(E) d\tilde x_0}
\end{pmatrix}
\end{eqnarray}
\end{widetext}
can be written in the form {$S_e(E)=e^{i(\theta+\bar k_F d)}D_\varphi(-\varphi_2)\tilde S_e(E)D_\varphi^\dagger(\varphi_2)$}, where $\tilde S_e(-E)=\tilde S_e^*(E)$. This allows us to rewrite the matrix $M$
in the form
{
\begin{equation}
\!\!\!M=  D_\varphi(\varphi_2) r_A(-\phi)\tilde S_e^T(E)r_A(\phi)\tilde S_e(E)D_\varphi^\dagger(\varphi_2).
\end{equation}
With this, we can then cast the equation $\alpha^2MW=W$ in the form $m\tilde W=m^*\tilde W$ with the matrix
\begin{equation}
m=\alpha r_A(\phi/2)\tilde S_e(E)r_A(-\phi/2).
\end{equation}}
Since the columns $\tilde {\cal A}$ of $\tilde W$ are either purely real or purely imaginary, this shows that $m\tilde {\cal A}$ is either purely real or purely imaginary. As a consequence the global phase of {$(\alpha S_{(n)}{\cal A}_{(n)})_k=(e^{i(\theta+\bar k_F d)}r_A(-\phi/2)D_\varphi(-\varphi_2))_{kk}(m_{(n)}\tilde A_{(n)})_k$} does not depend on energy, and $(\alpha S_{(n)}{\cal A}_{(n)})^*_k(\alpha S_{(n')}{\cal A}_{(n')})_k$ is real. 

With this we conclude that the phase of the current matrix operator elements is determined by the phase of $f_{nn',k}^\pm=e^{i\lambda_k(\epsilon_{n'}-\epsilon_n)({1\pm\tilde{x}'})/{2}}+\alpha_n \alpha_{n'}^* e^{-i\lambda_k (\epsilon_{n'}-\epsilon_n)({1\pm\tilde{x}'})/{2}}$, which is given as
\begin{align}
   \theta_{nn'}=\left(\arccos\epsilon_{n'}-\arccos\epsilon_n\right)/2.
\end{align}
As observed in Fig.~\ref{Fig::Figure_10}, the current operator matrix elements may vanish for particular values of $\tilde x$. In Fig.~\ref{Fig..x_dependency_matrix_elements}, we show that this is generically the case for all current operator matrix elements, both spin-preserving and spin-flip.

\onecolumngrid

\begin{figure}[ht!]
   \centering
   \includegraphics[width=1\linewidth]{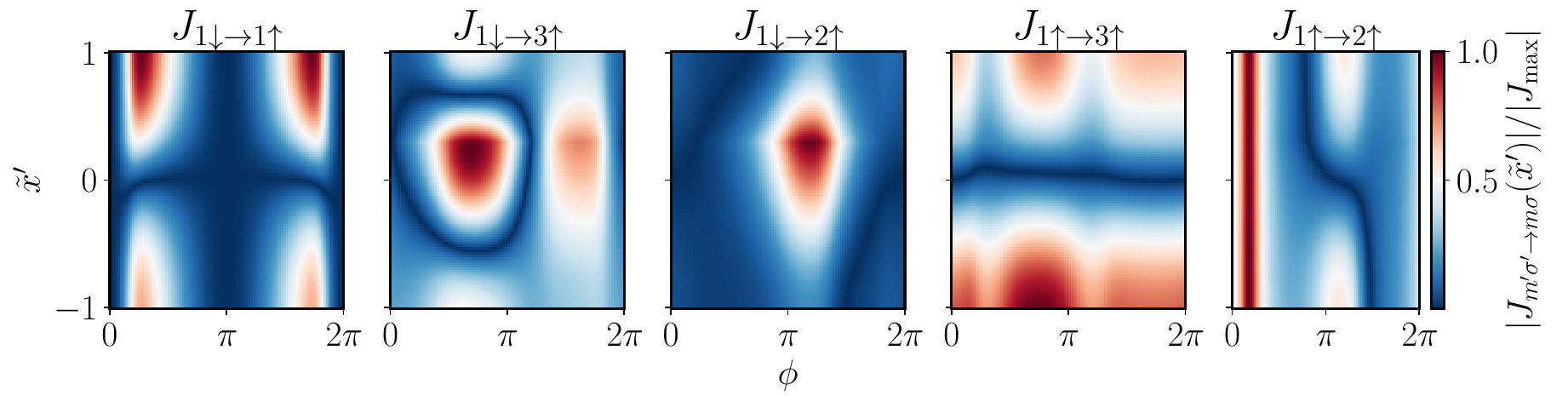}
   \caption{Dependence on the position of the phase drop $\tilde x'$ of several current operator matrix elements, covering all possible kinds of transitions. Same parameters as in Fig.~\ref{Fig::Figure_10}(d). We can see that both spin-preserving and spin-flip matrix elements generically vanish for particular values of $\tilde x$.}
   \label{Fig..x_dependency_matrix_elements}
\end{figure} }
\twocolumngrid

\bibliography{Bibliography}
\end{document}